\documentclass[useAMS,usenatbib]{mnras}
\usepackage{latexsym,amssymb,amsmath,amsfonts}
\usepackage{ae,aecompl} 
\usepackage{pslatex} 
\usepackage{rotfloat} 
\usepackage{rotating} 
\usepackage{float} 
\usepackage{graphicx}
\usepackage{longtable}
\usepackage{url}
\usepackage{dcolumn}
\usepackage{placeins}
\usepackage{multirow} 
\usepackage{hyperref}
\usepackage{rotating} 
\usepackage{ragged2e}
\justifying
\usepackage[T1]{fontenc}
\usepackage[none]{hyphenat}
\usepackage{color,soul}

\usepackage{multicol}
\usepackage{caption}
\usepackage{subcaption}
\usepackage[english]{babel}
\usepackage{blindtext}

\begin{document}
\newcommand{\sqcm}{cm$^{-2}$}  
\newcommand{\lya}{Ly$\alpha$}
\newcommand{\lyb}{Ly$\beta$}
\newcommand{\lyg}{Ly$\gamma$}
\newcommand{\lyd}{Ly$\delta$}
\newcommand{\hi}{\mbox{\tiny H\,{\sc i}}}
\newcommand{\HI}{\mbox{H\,{\sc i}}}
\newcommand{\HII}{\mbox{H\,{\sc ii}}}  
\newcommand{\Htot}{\mbox{{\rm H}}}

\newcommand{\HeI}{\mbox{He\,{\sc i}}}
\newcommand{\HeII}{\mbox{He\,{\sc ii}}}
\newcommand{\HeIII}{\mbox{He\,{\sc iii}}}  
\newcommand{\OI}{\mbox{O\,{\sc i}}}
\newcommand{\OII}{\mbox{O\,{\sc ii}}}
\newcommand{\OIII}{\mbox{O\,{\sc iii}}}
\newcommand{\OIV}{\mbox{O\,{\sc iv}}}
\newcommand{\OV}{\mbox{O\,{\sc v}}}
\newcommand{\OVI}{\mbox{O\,{\sc vi}}}
\newcommand{\OVII}{\mbox{O\,{\sc vii}}}
\newcommand{\OVIII}{\mbox{O\,{\sc viii}}} 
\newcommand{\CI}{\mbox{C\,{\sc i}}}
\newcommand{\CII}{\mbox{C\,{\sc ii}}}
\newcommand{\cstar}{\mbox{C\,{\sc ii}$^{*}$}}
\newcommand{\CIII}{\mbox{C\,{\sc iii}}}
\newcommand{\CIV}{\mbox{C\,{\sc iv}}}
\newcommand{\CV}{\mbox{C\,{\sc v}}}
\newcommand{\CVI}{\mbox{C\,{\sc vi}}}  
\newcommand{\SiII}{\mbox{Si\,{\sc ii}}}
\newcommand{\SiIII}{\mbox{Si\,{\sc iii}}}
\newcommand{\SiIV}{\mbox{Si\,{\sc iv}}}
\newcommand{\SiXII}{\mbox{Si\,{\sc xii}}}   
\newcommand{\AlII}{\mbox{Al\,{\sc ii}}}
\newcommand{\NiII}{\mbox{Ni\,{\sc ii}}}
\newcommand{\CrII}{\mbox{Cr\,{\sc ii}}}
\newcommand{\ZnII}{\mbox{Zn\,{\sc ii}}}
\newcommand{\Cst}{\mbox{C\,{\sc ii}$^{*}$}}
\newcommand{\OIst}{\mbox{O\,{\sc i}$^{*}$}}
\newcommand{\OIIst}{\mbox{O\,{\sc ii}$^{*}$}}
\newcommand{\SiIIst}{\mbox{Si\,{\sc ii}$^{*}$}}
\newcommand{\SII}{\mbox{S\,{\sc ii}}}
\newcommand{\SIII}{\mbox{S\,{\sc iii}}}
\newcommand{\SIV}{\mbox{S\,{\sc iv}}}
\newcommand{\SV}{\mbox{S\,{\sc v}}}
\newcommand{\SVI}{\mbox{S\,{\sc vi}}}  
\newcommand{\NI}{\mbox{N\,{\sc i}}}   
\newcommand{\NII}{\mbox{N\,{\sc ii}}}   
\newcommand{\NIII}{\mbox{N\,{\sc iii}}}     
\newcommand{\NIV}{\mbox{N\,{\sc iv}}}   
\newcommand{\NV}{\mbox{N\,{\sc v}}}    
\newcommand{\PV}{\mbox{P\,{\sc v}}} 
\newcommand{\NeIV}{\mbox{Ne\,{\sc iv}}}   
\newcommand{\NeV}{\mbox{Ne\,{\sc v}}}   
\newcommand{\NeVI}{\mbox{Ne\,{\sc vi}}}   
\newcommand{\NeVII}{\mbox{Ne\,{\sc vii}}}   
\newcommand{\NeVIII}{\mbox{Ne\,{\sc viii}}}   
\newcommand{\neviii}{\mbox{\tiny Ne\,{\sc viii}}}   
\newcommand{\NeIX}{\mbox{Ne\,{\sc ix}}}   
\newcommand{\NeX}{\mbox{Ne\,{\sc x}}} 
\newcommand{\MgI}{\mbox{Mg\,{\sc i}}}
\newcommand{\MgII}{\mbox{Mg\,{\sc ii}}}  
\newcommand{\MgX}{\mbox{Mg\,{\sc x}}}   
\newcommand{\FeII}{\mbox{Fe\,{\sc ii}}}  
\newcommand{\FeIII}{\mbox{Fe\,{\sc iii}}}   
\newcommand{\NaI}{\mbox{Na\,{\sc i}}}  
\newcommand{\NaIX}{\mbox{Na\,{\sc ix}}}   
\newcommand{\ArVIII}{\mbox{Ar\,{\sc viii}}}   
\newcommand{\AlXI}{\mbox{Al\,{\sc xi}}}   
\newcommand{\CaII}{\mbox{Ca\,{\sc ii}}}  
\newcommand{\TiII}{\mbox{Ti\,{\sc ii}}}
\newcommand{\ArI}{\mbox{Ar\,{\sc i}}}
\newcommand{\zabs}{$z_{\rm abs}$}
\newcommand{\zmin}{$z_{\rm min}$}
\newcommand{\zmax}{$z_{\rm max}$}
\newcommand{\zqso}{$z_{\rm QSO}$}
\newcommand{\zgal}{$z_{\rm gal}$}
\newcommand{\degree}{\ensuremath{^\circ}}
\newcommand{\lapp}{\mbox{\raisebox{-0.3em}{$\stackrel{\textstyle <}{\sim}$}}}
\newcommand{\gapp}{\mbox{\raisebox{-0.3em}{$\stackrel{\textstyle >}{\sim}$}}}
\newcommand{\be}{\begin{equation}}
\newcommand{\en}{\end{equation}}
\newcommand{\di}{\displaystyle}
\def\tworule{\noalign{\medskip\hrule\smallskip\hrule\medskip}} 
\def\onerule{\noalign{\medskip\hrule\medskip}} 
\def\bl{\par\vskip 12pt\noindent}
\def\bll{\par\vskip 24pt\noindent}
\def\blll{\par\vskip 36pt\noindent}
\def\rot{\mathop{\rm rot}\nolimits}
\def\alf{$\alpha$}
\def\lam{$\lambda$}
\def\refff{\leftskip20pt\parindent-20pt\parskip4pt}
\def\kms{km~s$^{-1}$}
\def\zem{$z_{\rm em}$}
\def\vrel{$v_{\rm rel}$}
\def\cmsq{cm$^{-2}$}
\def\cmcb{cm$^{-3}$}
\def\etal{et~al.\ }
\newcommand{\CLOUDY}{\mbox{\scriptsize{CLOUDY}}}
\title[DLA galaxies at $z\sim 3$]{Detection of emission lines from $z\sim3$ DLAs towards the QSO J2358+0149
\thanks{Based on observations carried out at the European Southern
Observatory (ESO) under programme ID 293.A-5020(A) (P.I. Noterdaeme) using X-Shooter spectrograph located on Unit Telescope 3 (UT3, Melipal) of the Very Large Telescope (VLT) at the Paranal Observatory, Chile.}}
\author[Srianand et al.]
{\parbox[t]{\textwidth}{Raghunathan Srianand$^1$\thanks{E-mail: anand@iucaa.in}, Tanvir Hussain$^1$, Pasquier Noterdaeme$^2$, Patrick Petitjean$^2$, 
 Thomas Kr{\"u}hler$^{3,4}$, Jure Japelj$^{5,6}$, Isabelle P\^aris$^6$ \&  Nobunari Kashikawa$^7$\\}
\\
$^{1}$ IUCAA, Postbag 4, Ganeshkhind, Pune 411007, India \\
$^{2}$ Institut d'Astrophysique de Paris, CNRS-UPMC, UMR 7095, 98bis bd Arago, 75014, Paris, France \\
$^{3}$ European Southern Observatory, Alonso de C{\'o}rdova 3107, Vitacura, Casilla 19001, Santiago 19, Chile\\
$^{4}$ Max-Planck-Institut f\"{u}r extraterrestrische Physik, Giessenbachstra\ss e, 85748 Garching, Germany\\
$^{5}$ Faculty of Mathematics and Physics, University of Ljubljana, Jadranska ulica 19, 1000 Ljubljana, Slovenia\\
$^{6}$ INAF-Osservatorio Astronomico di Trieste, via G. B. Tiepolo 11, 34131 Trieste, Italy\\
$^{7}$ Optical and Infrared Astronomy Division, National Astronomical Observatory, Mitaka, Tokyo 181-8588, Japan
}

\date{Accepted 2016 April 20. Received 2016 April 20; in original form 2015 October 13}

\label{firstpage}
\pagerange{\pageref{firstpage}--\pageref{lastpage}} \pubyear{2016}

\maketitle

\begin{abstract}
Using VLT/X-shooter we searched for emission line galaxies associated to four damped 
Lyman-$\alpha$ systems (DLAs) and one sub-DLA at 2.73$\le z\le$3.25 towards 
QSO J2358+0149. We detect [\OIII] emission from a ``low-cool'' DLA at \zabs = 2.9791 (having
log~$N$(\HI) = $21.69\pm0.10$, [Zn/H] = $-1.83\pm0.18$) at an impact parameter of, $\rho\sim 12$ kpc.
The associated galaxy is compact with a dynamical mass of $(1-6)\times 10^9$ M$_\odot$,
very high excitation ([\OIII]/[\OII] and [\OIII]/[H$\beta$] both greater than 10), 
12+[O/H]$\le$8.5 and moderate star formation rate (SFR$\le 2$ M$_\odot$ yr$^{-1}$). 
Such properties are
typically seen in the low-$z$
extreme blue compact dwarf galaxies. The kinematics of the gas is inconsistent with
that of an extended disk and the gas is part of either a large scale wind or cold accretion.
We detect \lya\ emission from 
the \zabs = 3.2477 DLA  (having log~$N$(\HI)=21.12$\pm$0.10 and
[Zn/H]=$-0.97\pm0.13$).
The \lya\ emission is redshifted with respect to the metal absorption lines by 320 \kms,
consistent with the location of the red hump expected in radiative transport models.
We derive SFR$\sim$0.2-1.7 M$_\odot$ yr$^{-1}$ and \lya\ escape fraction
of $\ge$10 per cent. No other emission line is detected from this system. Because the DLA has a small velocity separation from the quasar ($\sim$500 \kms) and the DLA emission is located within a small projected distance ($\rho<5$ kpc),
we also explore the possibility that
the \lya\ emission is being induced by the QSO itself. QSO induced \lya\ fluorescence is
possible if the DLA is within a physical separation of 340 kpc to the QSO. Detection of
stellar continuum light and/or the oxygen emission lines would disfavor this possibility.
We do not 
detect any emission line from the remaining three systems.  
\end{abstract}



\section{Introduction}

\label{intro}
Damped \lya\ systems (DLAs with $N(\HI) \geqslant 2 \times 10^{20} \rm cm^{-2}$), seen in the 
spectra of
bright background sources like QSOs, in principle trace galaxies based on their projected \HI 
gas cross-section and thereby providing a luminosity independent probe of high-$z$ galaxies
\citep[see][]{Wolfe05}. Thanks to large spectroscopic surveys, we now
have a very good statistical sample of DLAs in the redshift range $2\le z\le 4$ 
\citep{Prochaska05,Noterdaeme09dla,Noterdaeme12dla}.
DLAs at these redshifts contain $\sim$ 80\% of the neutral gas and the cosmic density of \HI\ gas in DLAs (i.e. $\Omega_{\rm H~{\sc I}}$)
shows an increasing trend at $2\le z\le 4$, while within statistical uncertainties evolving very mildly at $z\le 2$
At $z > 4$ the data does not show a clear trend either due to small number statistics or, especially when spectral resolution is low, from systematics due to blending with \lya\ forest
\citep[][]{Peroux03,Guimaraes09,Songaila10,Zafar13,Crighton15}.

Based on novel broad-band color selection and narrow-band imaging techniques, galaxies are detected up to $z\sim10$ 
and the star formation rate density of galaxies is mapped up to $z\sim10$ \citep[see][]{Madau14}. 
There is a strong increase in the UV luminosity density by 0.8 to 1 dex between $z=0$ and $z\sim2$ followed by a decrease 
towards higher $z$ for $z\ge3$. High spatial resolution deep imaging studies show, galaxies at $z\sim3$ have sizes typically
a factor 3 smaller compared to their local counterparts of similar luminosity \citep[e.g.,][]{Trujillo06}.

The observed number density of DLAs requires that the total \HI\ cross-section must be much larger than the optical extent of galaxies currently detected in emission, indicating wide spread of \HI\ around these galaxies (either in extended discs or in flows) or that a significant contribution to the
  gas cross-section comes from the population of faint galaxies below the detection limit of large surveys.
It is believed  that
the redshift evolution of the star formation rate density in the universe is driven by 
cold inflows \citep[see for example,][]{Keres05, Erb06,Ocvirk08,Dekel09}
and probably controlled by the large scale outflows that are ubiquitously found among the high redshift 
Lyman Break Galaxies 
\citep[LBGs,][]{Pettini01,Shapley03,Veilleux05}. These processes may control the physical conditions, 
chemical composition and kinematics of the extended gas 
distribution around galaxies (the so called circumgalactic medium) that are probed by absorption line studies. 
\begin{table*}
\caption{Log of X-shooter observations of J2358+0149$^+$}
\begin{tabular}{ccccccc}
\hline
\hline
PA & Observing date  & DIMM seeing & Air mass & \multicolumn{3}{c}{Exposure time(s)}\\
(deg) & dd/mm/year   &  (arcsec)   &          & UVB & VIS & NIR\\  
\hline
 0   & 02-07-2014&0.9 &1.14 & 2$\times$2900 & 2$\times$2800 & 2$\times$2880 \\
$+60$& 04-08-2014&1.0 &1.20 & 2$\times$2900 & 2$\times$2800 & 2$\times$2880 \\
$-60$& 01-07-2014&0.8 &1.20 & 1$\times$2900 & 1$\times$2800 & 1$\times$2880 \\
\hline
\end{tabular}\\
\begin{flushleft}
$^+$ Slit widths of 1.2, 0.8 and 1.3 arcsec were used for UBV, VIS and NIR 
spectrographs
of the X-shooter, respectively.
\end{flushleft}
\label{tablog}
\end{table*}

Purely based on the high resolution spectroscopic studies of DLAs we know that: (i) at a given $z$, the average metallicity 
of DLAs is typically less than those inferred for galaxies from their nebular emission, (ii) metallicity in DLAs shows clear 
redshift evolution but with a slope shallower than what one would have expected purely based on the galaxy 
evolution \citep[][]{Som13,Rafelski14} (iii) the volume filling factor of
cold gas (inferred through tracers like H$_2$ and 21-cm absorption) is lower than that seen in the local interstellar medium 
\citep[ISM,][]{Petitjean00,Ledoux03,Noterdaeme08,Srianand12,Kanekar14}
and (iv) DLAs show clear correlation between the velocity width of the low-ion absorption and the
metallicity akin to the mass-metallicity relation
seen in galaxies \citep[][]{Ledoux06a}. \citet{Moller13} explained the slow metallicity evolution as a
consequence of the lack of strong evolution in the zero-point of the mass-metallicity
relation of high-$z$ DLA galaxies.
\citet{Wolfe08} based on the cooling rate derived using \Cst\ absorption suggested the existence of
``low-cool'' and ``high-cool'' population of DLAs. They argued that both populations need local sources to satisfy the 
heating requirements \citep[see however,][]{Srianand05}. The rare detections of faint, extended objects in the Hubble Ultra Deep Field is incompatible with 
all heating coming from in situ star formation in ``high-cool'' DLAs and suggests that they may be originating 
from extended regions of LBGs \citep[][]{Wolfe06}. All this, while suggesting that the DLAs are somehow related to the star forming regions, also suggest
that they need not always be related to the gas in the stellar disk of galaxies. 

Therefore, identifying galaxies and correlating galaxy properties derived using continuum and nebular emission lines
to the properties of the associated DLAs derived from the high resolution spectra of QSO is very important. There have been
several attempts to detected galaxies associated to high-$z$ DLAs. 
However, searches aimed at directly detecting associated galaxies either in the continuum  emission 
or in the nebular line emission have mostly resulted in non-detections 
\citep[][as well as several unpublished works.]{Bunker99,Kulkarni00,Lowenthal95, Christensen09,Fumagalli15}
with very few cases being spectroscopically confirmed by the detection of \lya\ and/or other emission 
lines \citep[][]{Moller02, Moller04,Fynbo10,Noterdaeme12,Fynbo13,Krogager13,Bouche13, Hartoog15}.
These systems are used to obtain different correlations between properties derived from absorption
  lines towards quasars and emission lines from the associated galaxies. \citet{Krogager12} found a possible
  anti-correlation between $N$(\HI) and impact parameter (see their Fig.~3) similar to
  the one seen at low-$z$ based on \HI\ emission maps \citep[see Fig.~15 of][] {Zwaan05}.
  A possible correlation is also seen between the metallicity of the DLAs and the impact parameter.
  These correlations are shown to be consistent with a scenario where the metallicity-size
  relation is driven by starburst induced feed back processes. \citet{Christensen14} measured stellar masses of these DLA host galaxies. These are found to be consistent with the expected values based on
  mass-metallicity relations. As these interesting correlations are based on a handful of known
  systems studied in detail, it is important to increase the number of galaxies associated with DLAs.

Information on average star formation in DLA galaxies can in principle be derived using \lya\ stacking methods
\citep[][]{Rahmani10,Noterdaeme14}. However, complexities involved in the radiative transport of \lya\ photons
may hinder our interpretation in the absence of independent escape fraction measurements.

One possible way to enhance the efficiency of searching for DLA galaxies is to target QSO sightlines with
multiple DLAs in the optical near-IR regions. By scanning the full sample of QSO spectra from the SDSS-III
Baryonic Oscillation Spectroscopic Survey (BOSS) - Data Release 13 \citep[][]{Paris14} we have identified a QSO
sightline having 5 systems with \lya\ showing damping wings along the line of sight. 
In this paper, we present analysis of these five intervening DLA systems at \zabs = 2.7377, 2.9791, 3.1095, 3.1333 and 
3.2477 towards the SDSS J235854.4+014955.5 (hereafter refer to as J2358+0149). In particular, we report detection of [\OIII] and Ly$\alpha$ emission associated with DLAs at \zabs = 2.9791 and 3.2477 respectively. In Section~\ref{obs}, 
we provide details of the VLT/X-shooter observations, data reduction and systemic redshift measurements of the QSO. 
In Section~\ref{analyse} we present the analysis of the intervening absorption systems and a discussion on 
chemical abundances and dust depletions. We discuss, in Section~\ref{dgal1} and \ref{dgal2}, the emission properties 
of the DLA galaxies.
The summary and conclusions are presented in Section~\ref{conc}. In this paper 
we use a standard $\Lambda$CDM cosmology with $\Omega_\Lambda=0.73$, $\Omega_{m}=0.27$, and $\rm H_{0}=70~\rm km~s^{-1} \rm Mpc^{-1}$.

\section{Data and details of X-shooter observations:}
\label{obs}

Spectra of J2358+0149 were obtained  with X-shooter \citep{Vernet11} at the 
European Southern Observatory (ESO) Very Large Telescope (VLT) in 
service mode under Director's Discretionary Time (PI: Noterdaeme).  
Our observations consist of 5 observation blocks (OBs) for a total 
science exposure time of 14500 s in the ultra-violet-blue (UVB),
14000~s in the visible (VIS) and 14400~s in the near-infrared (NIR) arm
respectively. 
Each OB was performed using one nodding cycle (AB) with a nod throw of
4". NIR exposures were split into 480~s DITs (detector integration times).  
In order to be able to use the triangulation technique to get the galaxy impact parameter
with respect to the QSO line of sight using emission line centroids, we have
arranged the 5 OBs in
the following way:  two at position angles of 0$^\circ$ (North of East), two 
with PA=60$^\circ$ and one at position $-60^\circ$. Apart from
the differences in the position angle, other settings used in all OBs were identical. 
The observations were performed under good seeing (0.8-1.0 arcsec) conditions.
Detailed observational log is provided in Table~\ref{tablog}.

We reduced the X-shooter spectra using the ESO pipeline  \citep[version 2.5.2;][]{Goldoni06}. We performed
flat-fielding, order tracing, rectification and initial wavelength and flux calibration. For sky-subtraction, stacking of 
individual exposures, bad-pixel masking and cosmic ray rejection, we applied our own software and followed the procedure 
described in detail in \citet{Kruhler15}. Very briefly, we combined the individual, sky-subtracted exposures taken in a single
position angle with a weighted average, where the weight function was derived from the signal-to-noise ratio of the QSO spectrum. 
Bad pixel and cosmic rays were detected using a Laplacian edge detection filter and masked in the final stack. 
Error spectra, initially derived through the ESO pipeline, were propagated accordingly.

The X-shooter spectrum covers the wavelength range from 300~nm to 2.5~$\mu$m at 
the intermediate spectral resolution (R$\sim$6000 to 7000), thereby allowing  an
accurate measurement of the systemic redshift of the QSO and simultaneous search for the
\lya\ emission in the
blue and standard optical nebular lines like [\OII], [\OIII] and H$\beta$ in the NIR from the $z\sim3$ DLAs. 

\begin{figure}
\centering
  \includegraphics[bb= 20 360 560 730,width=1.\linewidth,clip=true]{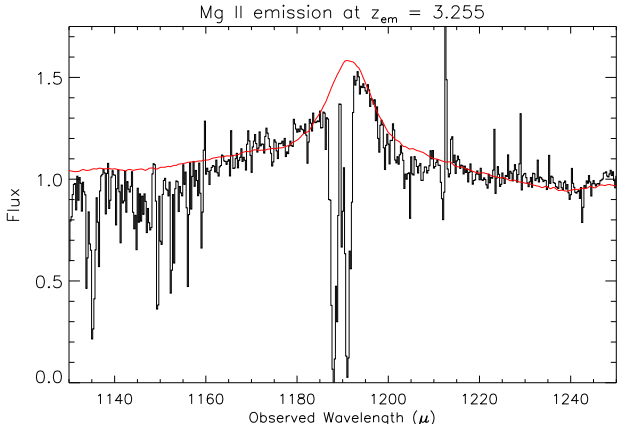}
  \caption{The Mg~{\sc ii} broad emission line detected in the near-IR
spectrum of J2358+0149. Composite QSO spectrum redshift to \zem = 3.255 is overlayed
for comparison.} 
  \label{tem}
  \end{figure}
In the optical UV spectrum of the QSO the \lya\ + \NV\ emission lines are severely absorbed by the associated DLA
absorption at \zabs = 3.2477.
The [\CIII], \SiIV\ and \CIV\ emission lines are also week, broad and asymmetric compared to that of the
QSO composite spectrum. Based on \CIV\ and \SiIV\ emission lines we get the QSO
redshift of \zem = 3.235.  We also do not detect strong [\OIII] emission from the QSO though there
is a weak feature coinciding with the H$\beta$ emission.
The only prominent  broad emission line that is symmetric is \MgII\ line
that we detect in the near-IR spectrum (see Fig~\ref{tem}). We fit the \MgII\ profile with a Gaussian  and estimated 
the emission redshift to be \zem = 3.255$\pm$0.001. In what follows, we use \zem = 3.255 for all practical purposes.

\begin{figure*} 
\centering{
\includegraphics[bb=180 25 580 740,width=0.55\linewidth,angle=270,clip=true]{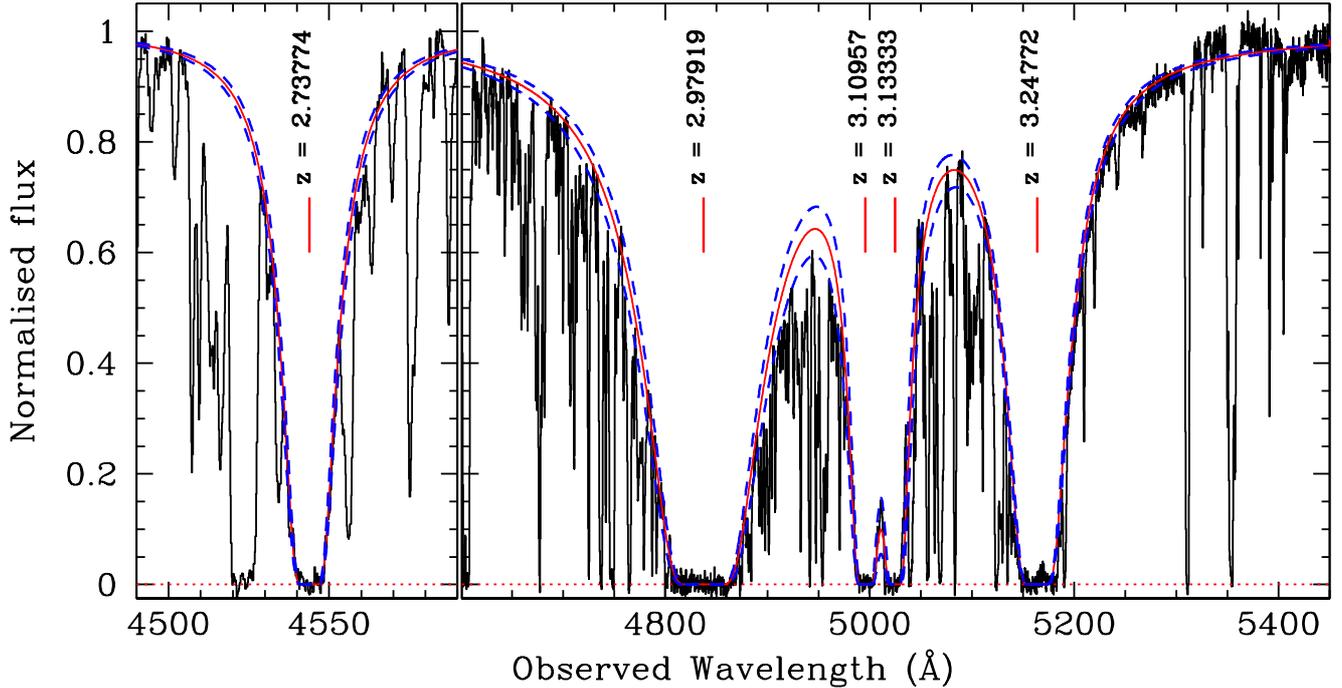} 
}
\caption{{\it Left panel:} Voigt profile fits to the \lya\ absorption from the sub-DLA at \zabs = 2.7277.
{\it Right panel:} Simultaneous Voigt profile fit to  \lya\ absorption from four DLAs at \zabs = 2.9791, 3.1095, 3.1333 and
3.2477. Dashed profiles show the 1$\sigma$ range.}          
\label{dla_hi}
\end{figure*}  

\section{Brief description of metals in individual systems}
\label{analyse}

Along the J2358+0149 sightline, we detect five absorption systems with \lya\ absorption showing damping wings : one sub-DLA (at \zabs = 2.73773)  and four DLAs (at \zabs = 2.97919, 3.10956, 3.13333 and 3.24772). We estimate the column densities of
\HI\ and metal ions using the Voigt profile fitting code
{\sc vpfit}\footnote{\url{http://www.ast.cam.ac.uk/~rfc/vpfit.html}} version 10.0. The Voigt profile fits to the \lya\ absorption lines
are shown in Fig~\ref{dla_hi}. In the case of metal lines, while fitting the profiles\footnote{As the seeing during our
observations was better than the slit width used, the actual resolution achieved is slightly better than the normal value.
Therefore, we adjusted the instrumental profile by fitting the narrow telluric absorption lines. The measured resolution is $R \approx 9700$ and $6000$ in the visible and UVB respectively. We use these values to generate instrumental profiles while fitting the absorption lines. Note this effect is not important in the case of extended line emission from galaxies.} we keep the $z$ and $b$-parameter 
for each component to be same for all the species
(i.e. species are assumed to be tied to each other and the absorption profile is dominated by non-thermal motions).
Due to the intermediate resolution of the X-shooter spectrum, hidden saturation in the narrow lines can hinder our metallicity
measurements. Therefore, as far as possible we discuss the metallicities of species that were measured using several
lines having a wide range of oscillator strengths. In particular this is true for \FeII\ and \SiII. Voigt profile 
fits overlayed on the data for all the 5 systems are presented in the Appendix (see Fig.~\ref{vpfit1}-\ref{vpfit5}).  We also summarize the
  results of Voigt profile fitting of individual components in tables given in the Appendix (see tables~\ref{at1}-\ref{at5}). The set of
  transitions used to derive these fits are provided when we discuss individual systems below. The total column densities of different species obtained are summarized in Table~\ref{tab1}.
The metallicity\footnote{[Z/H]~$~\equiv~\log$(Z/H)$_{\rm obs} - \log$(Z/H)$_{\odot}$} of individual ions detected (without applying
ionization corrections) in  these five systems are summarized in Table~\ref{tab2}. We provide a short summary of each system in this section.
For all discussion on metallicity we use solar relative abundances of the metals taken from \citet{Asplund09}.

\begin{table*}
\centering
\caption{Total column densities of metals and neutral hydrogen in DLAs}
\label{tab1}
\resizebox{\textwidth}{!}{%
\begin{tabular}{lccccccc}
\hline
\hline
\multicolumn{1}{c}{} & \multicolumn{7}{c}{{\Large\strut}log $N$ ({\sqcm})} \\ 
\cline{2-8}
     \zabs         & \HI                         & \SiII$^{1}$        & \SII            & \FeII          & \NiII          & \CrII          & \ZnII          \\ 
\hline
2.73774          & $20.07\pm0.05$        & $13.90\pm0.14$ & $ < 13.73\pm0.35$  & $13.52\pm0.33$ &   $ - $           &  $ -$             &  $- $             \\ 
2.97919          & $21.69\pm0.10$        & $15.49\pm0.24$ &   $-$             & $15.21\pm0.15$ & $13.70\pm0.31$ & $13.16\pm0.21$ & $12.42\pm0.15$ \\
3.10957          & $20.45\pm0.10$        & $14.66\pm0.09$ &    $-$            & $14.37\pm0.11$ &   $ -$            &    $- $           &   $- $            \\ 
3.13333          & $20.33\pm0.10$        & $14.67\pm0.45$ &     $-$           & $13.82\pm0.16$ &   $-$             &  $-$              &  $ - $            \\ 
3.24772          & $21.12\pm0.10$         & $15.68\pm0.12$ & $15.08\pm0.08$  & $15.12\pm0.12$ & $13.64\pm0.17$ & $13.39\pm0.09$ & $12.71\pm0.09$   \\ \hline
\end{tabular}
}
\begin{flushleft}
$^{1}$ The quoted column density can be affected by the presence of hidden saturation components if they have $b$ < 4 \kms. See text.
\end{flushleft}
\end{table*}

\begin{table*}
\centering
\caption{Metal abundances in DLAs}
\label{tab2}
\resizebox{\textwidth}{!}{%
\begin{tabular}{lccccccc}
\hline
\hline
\zabs    & {[}Si/H{]}     & {[}S/H{]}           & {[}Fe/H{]}     & {[}Ni/H{]}     & {[}Cr/H{]}     & {[}Zn/H{]}     \\ 
\hline
2.73774 & $-1.68\pm0.14$ & $ < -1.81$      & $-2.05\pm0.33$ & $-$           & $-$             & $-$           \\
2.97919 & $-1.71\pm0.26$ & $-$             & $-1.98\pm0.18$ & $-2.21\pm0.32$ & $-2.17\pm0.23$ & $-1.83\pm0.18$ \\
3.10957 & $-1.30\pm0.13$ & $-$             & $-1.58\pm0.15$ & $-$              & $-$              & $-$              \\
3.13333 & $-1.17\pm0.46$ & $-$             & $-2.01\pm0.19$ & $-$              & $-$              & $-$              \\
3.24772 & $-0.95\pm0.15$ & $-1.16\pm0.13$  & $-1.50\pm0.16$ & $-1.70\pm0.20$ & $-1.37\pm0.13$ & $-0.97\pm0.13$ \\ 
\hline
\end{tabular}
}
\end{table*}

\subsection{The sub-DLA at \zabs = 2.7377} 

The Voigt profile fit to the damped \lya\ absorption gives log~$N$(\HI) = 20.07$\pm$0.05 (see Fig.~\ref{dla_hi}).
In addition to the \lya, absorption lines from \SiII, \SII, \FeII\  and \MgII\ lines are detected. At the resolution
of our spectrum these lines are unresolved. We measure rest equivalent width ($W_r$) of 0.50 and 0.40 \AA\ for the \MgII\
doublets. The expected position of \CII\ absorption falls in the \lya\ absorption of the \zabs = 2.9791 system.
Therefore, we could not have a handle on the gas cooling rate for this system.
To get a better constraint on the column density measurements, we perform Voigt profile fits of \SiII\ \lam1526 and 
\FeII\ \lam\lam\lam2382,2374,2344. The best fit Voigt profiles for this system is shown in 
Fig.~\ref{vpfit1} and measured column densities are shown in Table.~\ref{tab1}. A 3$\sigma$ upper limit to the column density 
is measured for blended \SII\ transitions in this system.  The metallicity of the gas based on \SiII\ absorption
is [Si/H] = $-1.68\pm0.14$. The measured upper limit on sulphur metallicity based on \SII\ (i.e. $\le-1.81$) are consistent with this metallicity 
(see Table~\ref{tab2}). 
Based on the observed Fe~{\sc ii} column density we derive [Fe/H] = $-2.05\pm0.33$. Even though the mean metallicity
of Fe suggests a possible under abundance, the errors are too large to draw any firm conclusion.
We do not detect H$_2$ from this system. Given the low metallicity and low $N$(\HI) measured in this system, 
non-detection of H$_2$ is not at all surprising \citep[][]{Petitjean06}.
This system does not show a strong \CIV\ absorption (with $W_r$(\CIV\ \lam1548) $\sim0.05$ \AA). However,
it is interesting to note that the nearest \lya\ absorption system with a velocity separation of $\sim 950$ \kms\ (i.e. at \zabs = 2.7259, the line seen at $\sim$4530~\AA\ in
Fig.~\ref{dla_hi}) shows a strong \CIV\ (with  $W_r$(\CIV\ \lam1548) $\sim0.34$ \AA) and \SiIV\ absorption.
In our 2D spectrum we do not detect any associated emission line galaxy at all three position angles within the 
impact parameter probed (i.e. $\le 15$ kpc).
  
\subsection{The DLA at \zabs = 2.9791}
\label{sect2}
This is the system with the highest \HI\ column density along the line of sight to J2358+0149. 
The Voigt profile fit to the \lya\ line gives log~$N$(\HI) = $21.69\pm0.10$ (see Fig.~\ref{dla_hi}).
This is close to the limiting column density (log~$N$(\HI) = $21.7$) used by \citet{Noterdaeme14} 
to define extremely strong DLAs (ESDLAs).
In this system, absorption lines of \CII, \SiII, \CrII, \FeII, \NiII\ and \ZnII\ along with \Cst\ are 
detected. Even at the spectral resolution of the X-shooter most of the  absorption lines are saturated.
In the NIR spectrum we detect very strong \MgII\ (with $W_r\sim$ 2.6~\AA\ and 2.2~\AA\ for the doublet) \FeII\ \lam2600 (with  $W_r\sim1.9$~\AA) and  \MgI\ 
(with $W_r\sim1.7$~\AA) lines. Thus purely based on \MgII\ absorption
strength this system is in the lower end of ``{\it ultra-strong}'' \MgII\ systems. 
Such absorbers show very large velocity
spread and are thought to be associated either with supernova driven winds \citep[see for example,][]{Nestor11} or 
cold gas embedded in intragroup media \citep[][]{Gauthier13}. Using the JHU-SDSS catalog \citep[][]{Zhu13} of \MgII\ absorbers
we find that only 0.5\% of the \MgII\ absorbers at \zabs$\ge$2.0 have such large equivalent widths for \MgII\ and
\FeII. Thus based on observed $N$(\HI) and equivalent widths of \MgII\ and \FeII\ the present system
seems to be a rare system among the QSO absorption line systems.

As expected the absorption line from singly ionized species spread over $\sim 240$ \kms\ whereas \CIV\ absorption
spread over 400 \kms (see Fig.~\ref{ema}). Using the definition of \citet{Ledoux06a}, we measure $\Delta v_{90}$\footnote{Using the moderate resolution spectrum can bias the derived $\Delta v_{90}$ values.
    However, for the spectral resolution of the X-shooter, this effect is expected to be small for the   $\Delta v_{90}$
    value we find here \citep[see Fig.~1 of][]{Arabsalmani15}.}
of 168, 141 and 144 \kms\ for \FeII\ \lam1608, \SiII\ \lam1808 and \CIV\ \lam1548 respectively.
Only $\sim 19\%$ of high-$z$ DLAs show such a 
large velocity spread of low ion absorption lines
\citep[see Fig.~9 of][]{Noterdaeme08}. The velocity spread similar to the one we see for \CIV\ is also very rare in 
DLAs \citep[see Fig.~2 of][]{Fox07a} and in intervening \lya\ forest absorbers \citep[see Fig.~11 of][]{Muzahid12}.
We will discuss the implications of the large velocity width in detail in Section~\ref{dgal1}.

The Voigt profile fits to the absorption lines were performed using \SiII\ \lam\lam1526,1808, \FeII\ \lam\lam\lam1608,2249,2260, 
\CrII\ \lam2062, \NiII\ \lam1709, \ZnII\ \lam2026 and \Cst\ \lam1335.7 simultaneously with five components. 
The contribution of the blends \CrII\ \lam2026 and \MgI\ \lam2026 in \ZnII\ \lam2026 has to be taken care 
  of while fitting. As we do not have independent constraints on \MgI\ column densities, uncertainties in the
  derived \ZnII\ column density may be higher than the statistical error we quote.
The derived parameters are presented in Table~\ref{tab1} and the 
fit is shown in Fig.~\ref{vpfit2}. We measure [Zn/H] = $-1.83\pm0.18$ and [Si/H] = $-1.71\pm0.26$.
This is consistent with a low metallicity gas without relative enhancement of $\alpha$-elements with respect
to the Fe co-production elements.
As can be seen from Fig.~\ref{vpfit2}, the absorption profile is well
fitted with 5 components with a reduced $\chi^2_\nu\sim 1$.
However, from Table~\ref{at2} we notice that the main component has a large
$b$-value. In order to check the possible hidden narrow saturated component in this velocity range, we fitted the absorption profile by fixing the
$b$ parameter in the range 3-8 \kms\ as typically derived for metal lines observed at high-resolution. Values lower than this are 
usually associated with H$_2$ components \citep{Ledoux03,Srianand05}. The derived column densities
are consistent (i.e. at most 0.1 dex higher) with our best fit values in the case of \CrII, \ZnII, \FeII\ and
  \NiII\ as we have good constraints from weak lines that have residual
  fluxes consistent with non-saturation even at these $b$-values. However,
  in the case of Si~{\sc ii} for $b$ = 3 \kms\ we could accommodate a factor of two higher
  column density in a hidden saturated component compared to our best fitted value. Therefore, our \SiII\ absorption is susceptible to undetected hidden saturation and high resolution spectrum is needed to get an accurate
  column density.

The dust content in this 
absorber can be  determined from the depletion factors of metals detected 
compared to zinc:  $\rm [Cr/Zn] = -0.34\pm0.29$, $\rm [Fe/Zn] = -0.15\pm0.25$ and  
$\rm [Ni/Zn] = -0.38\pm0.37$. The small depletion seen in this very high \HI\ column density absorber
is consistent with the lack of strong reddening signatures in the QSO continuum.
We do not detect the absorption from \CI, \TiII, \NaI\ and \CaII.
The expected position of fine-structure lines of \OI\ falls in the red wing of the DLA 
at \zabs= 3.2477. We do not detect fine-structure lines of \SiII\ also. We differ a detailed
discussions on \Cst\ cooling rate to Section~\ref{dgal1}.

It has been suggested that H$_2$ may be more frequently detected in ESDLAs compared to the
normal DLAs \citep[see][]{Noterdaeme15,Noterdaeme15a}. We do not detect H$_2$ in any of
the Voigt profile component of the low ions. We place a 3$\sigma$ upper limit on log~$N$(H$_2$, J=0)$\leq$ 15.54 (3$\sigma$)
and for log~$N$(H$_2$,J=1)$\leq$15.70 (3$\sigma$) for a line having a typical width of the narrowest line detected in the
\lya\ forest. We detect an associated galaxy through a strong [\OIII] emission line.
While the BOSS spectrum of the QSO does show an emission feature in the middle of the \lya\ absorption
we do not detect any \lya\ emission in our X-shooter spectrum.
Detailed analysis of the galaxy is presented in Section~\ref{dgal1}.

\subsection{The DLA at \zabs = 3.1095}

We measure  $\log N(\rm \HI) = 20.45 \pm 0.10$ for this system. At the X-shooter's resolution
the absorption is mostly concentrated in one component. Absorption from \ArI, \NI, \OI, \CII, \SiII, \FeII, \MgII\  and \MgI\ are observed in this system.
From the NIR spectrum we measure the equivalent widths of \MgII\ ($W_r\sim$ 1.00 and 0.88~\AA\ for the doublet), \FeII\ \lam2600 ($W_r\sim$ 0.58~\AA) and \MgI\ ($W_r\sim$ 0.18~\AA). We performed Voigt profile fits with two components for \NI\ \lam1199.5, \OI\ \lam\lam\lam971,976,1039, \ArI\ \lam1048,  
\CII\ \lam1334, \SiII\ \lam\lam1526,1808 and \FeII\ \lam\lam\lam1608,2344,2374 transitions. It is possible that our column density
estimations are affected by unresolved saturation in the strong resonance transitions like \CII.
The best fit Voigt profile to the ions is shown in Fig.~\ref{vpfit3}. We derive [Si/H]=$-1.30\pm0.13$, [O/H]=$-1.87\pm0.20$,
[N/H]=$-2.8\pm0.14$, [Ar/H]=$-1.61\pm0.48$ and [Fe/H]=$-1.58\pm0.15$. The large variations in the metallicity
measurements may be a consequence of hidden line saturation (the main component is narrow and requires low
$b$-value as can be seen in Table~\ref{vpfit3}) and one needs to be cautious in interpreting them.
The metallicity measurements are consistent with very little
or no depletion of Fe.  As typically seen in other DLAs we also notice [N/O] being much below the primary line \citep[see,][]{Petitjean08,Cooke11b,Dutta14}. The measured abundance pattern makes this system an interesting target for metallicity measurements with high resolution spectroscopy.
We do not detect any emission line galaxy associated to this DLA. 


\subsection{The DLA at \zabs = 3.1333}

The $\log N(\rm \HI) = 20.33 \pm 0.10$ measured in this system just about satisfies the definition of a DLA.
Absorption from \CII, \SiII, \FeII\ and \OI\ are observed in this system. The \MgII\ ($W_r$$\sim$0.49 and 0.44~\AA\ for the doublet) and \FeII\ \lam2600 ($W_r$$\sim$0.27~\AA) lines are
relatively weak compared to the other systems. 
The column density and $b$-parameter measurements 
of the observed ions are derived from \CII\ \lam1036, \SiII\ \lam\lam1526,1260, \FeII\ \lam\lam\lam2382,2374,2344 and \OI\ \lam1039. The Voigt profile fit to the metals and  measured column densities are shown in Fig.~\ref{vpfit4} and Table~\ref{tab1}. 
The derived [Si/H]=$- 1.17 \pm 0.46$ is higher than that of the other DLAs discussed above. We also find [Fe/Si]=$-0.82\pm0.50$.
As column densities of \FeII\ and \SiII\ were obtained using several transitions, this low value of Fe could reflect
dust depletion.  Due to the low measured $N$(\HI) we expect this absorber to produce little extinction even if there
is dust depletion at the level inferred above. Like the \zabs = 3.1095 system the measured \OI\ column
density (i.e. $\log N$(\OI)=14.93$\pm$0.26) suggests a lower [O/H]=$-2.09\pm0.28$ compared to that inferred from the 
Si measurements. As in the case of the previous system the $b$-value for the main component is low (see Table~\ref{vpfit4}) and hidden
saturation could affect our metallicity measurement, it will be interesting to measure these abundances with high resolution spectra.

\subsection{The DLA at \zabs = 3.2477}

The measured \zabs\ of this DLA is very close to the QSO redshift, and the measured velocity, $v = -515$ \kms, with
respect to the QSO is consistent with the system being 
a proximate DLA \citep[pDLA,][]{Ellison10}. We measure  $\log N(\rm \HI) = 21.12 \pm 0.10$ by fitting the DLA profile.
In this system, we detect absorption from \CII, \SII, \SiII, \FeII, \MgII, \MgI, \CrII, \NiII\ and \ZnII. The
equivalent width of \MgII\ absorption ($W_r\sim3.8$ and 4.0~\AA\ for the doublet) is much higher than what we have seen for
the \zabs = 2.9791 DLA. However \MgI\ absorption is relatively weaker (i.e. $W_r$(\MgI)$\sim$0.27~\AA) compared to that measured 
in the other systems. This may imply an extended velocity structure in the present system
compared to the \zabs = 2.9791 DLA.

Absorption profiles of strong low-ion transitions like \SiII~\lam1526 and \AlII~\lam1670 spread over $\sim 500$ \kms,
while those of relatively weak low-ion transitions like \FeII~\lam1608 spread over 250 \kms\  (see Fig.~\ref{ema1})
has $\Delta v_{90}$= 149 \kms. 
Unlike in the case of the \zabs = 2.9791 DLA, the high ionization phase probed by the \CIV\ absorption is not wider than that
of the low-ions. However, the \CIV\ absorption profile is distinctly different from those of the low-ions and spread over 
$\sim$ 250 \kms. Despite the pDLA being very close to the QSO we do not detect \NV\ or \OVI\ absorption associated
to the \CIV\ absorption \citep[as often seen in the associated absorbers in][]{Petitjean94a}.

 \vspace*{-0.04cm}
Detection of absorption from the excited fine-structure levels of \CII, \SiII\ and \OI, in the associated
absorption systems, can be used to infer the distance of the absorber from the QSO \citep[see for example,][]{Srianand00apm,Fathivavsari15}. Due to the large velocity spread, \Cst\ \lam1335 from the strong components are blended with \CII\ \lam1334
from the velocity components in the red. Similarly, the expected wavelength range of \OIst\ \lam1304 and \OIIst\ \lam1306 absorption falls with the profile of \SiII\ \lam1304 absorption.
High resolution spectra are needed to measure the column densities of these species. However, the expected wavelength
range of \SiIIst\ \lam1533 falls in the blending free region. We do not detect any clear absorption line. For the main
component we get log~$N$(\SiIIst)$\le12.30$ (3$\sigma$). 

Multi-component Voigt profile fit is performed using
\SiII\ \lam\lam1808,1526, \SII\ \lam1253, \FeII\ \lam\lam\lam1608,2249,2260, \CrII\ \lam2066, \NiII\ \lam\lam\lam1741,1709,1751 
and \ZnII\ \lam2026 transitions and is shown in Fig.~\ref{vpfit4}. 
Both the  \CII\ \lam1334,1036 transitions are heavily saturated and hence were not included while fitting.

The column density of \ZnII\ \lam2026 has negligible contribution from \CrII\ \lam2026 and 
\MgI\ \lam2026 lines. The metallicity for this DLA is $[\rm Zn/H] = -0.97\pm0.13$. Within measurement
uncertainties, this is consistent with [Si/H] = $-0.95\pm0.15$ and [S/H] = $-1.16\pm0.13$.
Therefore, for this systems also we do not find any indication of enhanced $\alpha$-element
abundance compared to the Fe co-production elements. We tested the effect of possible hidden saturation as explained in Sect.~\ref{sect2}. Similar to the \zabs = 2.9791, we find that the column densities of \ZnII,  \CrII, \FeII\  and \NiII\ are not dependent on the assumed $b$-values. However, for \SiII, the presence of possible hidden saturation could increase the observed column density by upto 1 dex if $b \leqslant 4$~\kms. While high-resolution is required to accurately measure $N$(\SiII), a one dex higher value would lead to a [Si/Zn] ratio much above the Solar value, which is unlikely.

The dust content in this DLA can be determined from the depletion
factors of iron, nickel, chromium, silicon compared to zinc. We measure,
[Cr/Zn] = $-$$0.40\pm0.18$, [Ni/Zn] = $-$$0.73\pm0.24$ and [Fe/Zn] = $-$$0.53\pm0.20$.
The moderate depletion noted here is consistent with what is seen  in
the halo gas in our galaxy. We also measure the dust to gas ratio ${\rm \kappa = 10^{[Zn/H]}(1.-10^{[Fe/Zn]})} = 10^{-1.12}$
and iron column density in dust of log~${\rm N_{Fe}^{Dust} = 15.48}$.
Systems having metallicity, $\kappa$ and ${\rm N_{Fe}^{Dust}}$ as measured in the present case are known to have
a high probability for H$_2$ detections \citep[see][]{Petitjean06,Noterdaeme08}. However we do not detect H$_2$ in 
any of the low-ion component and we place the limiting column densities of log~$N({\rm H_2,J=0}) \le 15.0$ (5$\sigma$)
and   log~$N({\rm H_2,J=1}) \le 15.2$ (5$\sigma$) for the strongest low-ion component.
We detect \lya\ emission associated to this DLA. However, no other emission line is clearly detected. We discuss 
this in detail in section~\ref{dgal2}.

\section{Properties of the galaxy associated to the \zabs = 2.9791 DLA}\label{dgal1}
\begin{figure*}
  \centering
 \vbox{ 
 {\includegraphics[bb = 70 10 600 750,height=1.0\linewidth,angle=90,clip=true]{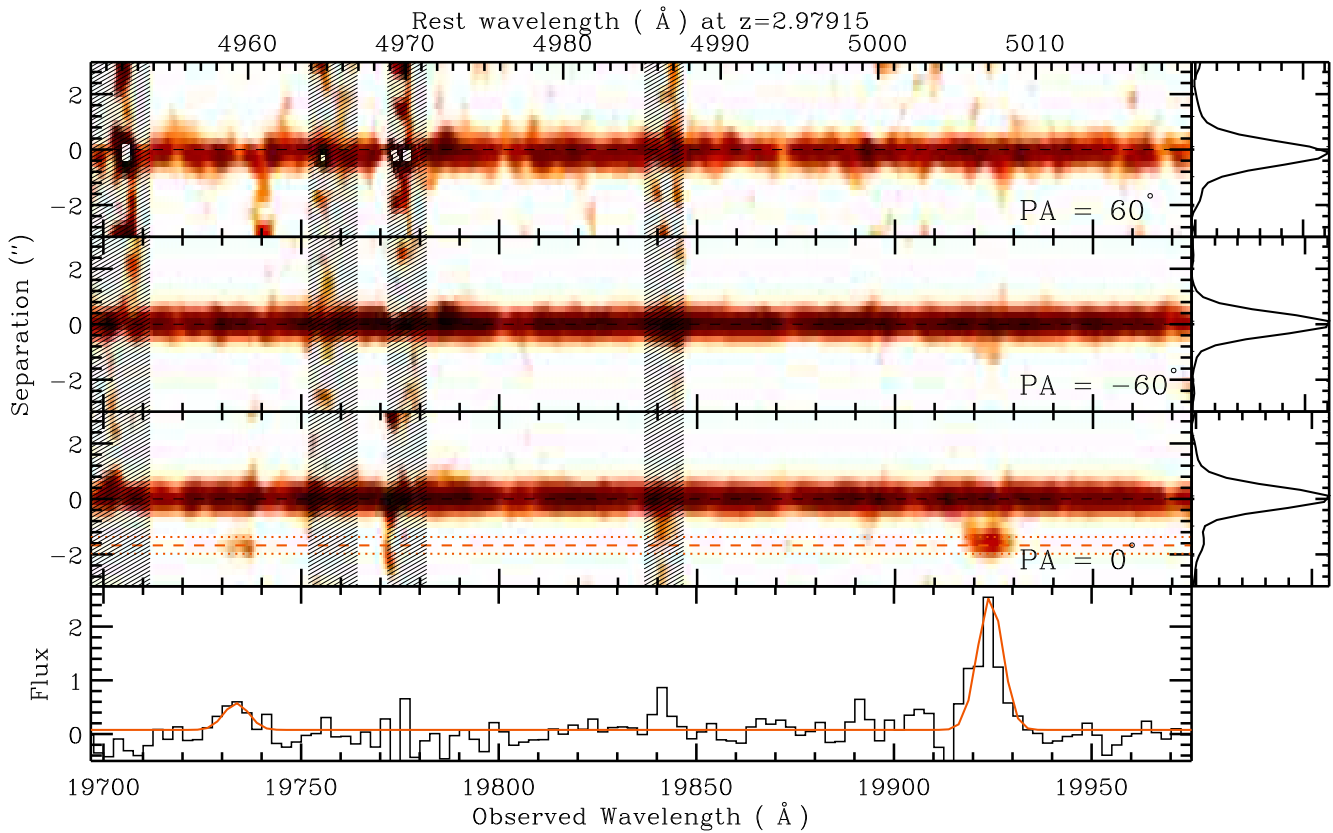}}
 }
 \caption{{\it Top three panels}: 2D spectrum of the J2358+0149 obtained with the slit oriented in
three position angles (mentioned in each panel) in the expected wavelength range of [\OIII] emission
at the redshift of the \zabs = 2.9791 DLA.The  [\OIII] emission, well detached from the QSO trace, is
seen only for PA = 0$^\circ$. The trace of the galaxy is shown with a long-dashed line and the dotted lines give the
$1\sigma$ of the trace. Shaded regions
mask the wavelength range affected by sky-subtraction residuals.  [\OIII] emission lines are not detected in the spectra
taken with other two position angles.
{\it Bottom panel:} 1D spectrum extracted at the galaxy trace is shown together with a single 
Gaussian fit to the [\OIII] lines. The flux is given in units of $10^{17}$ erg s$^{-1}$ cm$^{-2}$ \AA$^{-1}$.
} 
\label{oiii_lowz}
\end{figure*}

\begin{table}
 \centering
 \caption{Emission lines from the DLA galaxy at \zgal=2.9784}
 \label{tab4}
\begin{tabular}{cccc}
\hline \hline
Line & Flux & Luminosity \\
  & ($10^{-17}~\rm erg~cm^{-2}~s^{-1}$) & ($10^{42}~\rm erg~s^{-1}$)  \\ \hline
 \lya & $ \le$ 0.64$^a$  & $ \le 0.58 $   \\
$[$\OII$]$\lam3727   & $\le 0.19^b$ & $ \le 0.15 $ \\ 
$[$H$\beta]$\lam4862 & $ \le 0.22^b$ & $ \le 0.17 $   \\
$[$\OIII$]$\lam4960  & $0.52\pm0.06$ & $0.41\pm0.04$ \\ 
$[$\OIII$]$\lam5008  & $2.02\pm0.15$ & $1.59\pm0.12$\\ \hline 
\end{tabular}\\
\begin{flushleft}
$^a$ Obtained assuming \lya\ line spread as expected in the static medium (see text for details).
$^b$ Using the line width identical to that of \OIII.
\end{flushleft}
\end{table}

We detect [\OIII] \lam\lam 4960, 5008 emission in the spectrum obtained with the slit
aligned at a position angle (PA) = 0\degree\ for the DLA at \zabs = 2.97919 (See Fig.~\ref{oiii_lowz}). 
The emission
is seen at a projected separation of $\sim1.5\pm0.1$ arc sec from the QSO trace that 
corresponds to an impact parameter of $\rho =  11.9\pm0.8$ kpc. The fact that the emission is not
seen in spectra taken with slit oriented in other two position angles is consistent
with the extent of the emitting region being smaller than 13 kpc in the East-West 
direction. The emission is unresolved in the
spatial axis suggesting that the galaxy is compact with a size of $\le 4.5$ kpc. We do not
detect other emission lines like Ly$\alpha$, H$\beta$ and [\OII]~\lam3727 at the
spatial location where the [\OIII] emission is detected.
We derive the [\OIII] line flux by fitting Gaussians to the observed line profiles 
(see bottom panel of Fig.~\ref{oiii_lowz}).

\begin{figure}
  \centering
  \vbox{
  \includegraphics[bb=240 10 455 770,height=1.\linewidth,angle=90,clip=true]{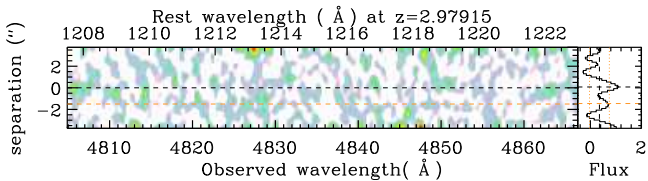}
}
  \caption{2D-spectrum at the expected wavelength range of \lya\ emission from the \zabs = 2.9791 DLA
obtained using the slit aligned at a PA of 0$^\circ$.
The original image is smoothed by a Gaussian filter having an FWHM of 3 pixels (i.e. 0.6 \AA) along the 
wavelength axis and 5 pixel (i.e. 0.75 arc sec) along the spatial axis. In the right hand side panel,
the average flux (in units of $10^{-19}$ erg s$^{-1}$ cm$^{-2}$ \AA$^{-1}$) is shown for each row. Based
on the [\OIII] emission detected we expect the \lya\ emission at the spatial separation of 1.5 arcsec
from the QSO trace (shown at 0 along the y-axis). }
  \label{lyaz2p97}
\end{figure}

In the case of other non-detected emission lines, we derive the 3$\sigma$ upper limits by assuming 
the line width to be similar to that of the [\OIII] line. As the \lya\ emission can be much wider than the
rest of the lines, due to radiative transport effects, we follow a different procedure to get its
limiting flux. As can be seen in Fig.~\ref{lyaz2p97}, no significant flux is detected along the 
expected trace of  the emission line galaxy.  
If we assume 
the gas to be a static slab, then the \lya\ profile will have double humps separated
by $\sim$1200 (respectively 526) \kms\ if we assume the gas temperature to be 10$^4$ K (respectively
for 100 K) for the measured $N$(\HI) \citep[using eq. 21 of][]{Dijkstra14}. We integrated 
the flux within $\pm$ 600 \kms\ to the redshift of the [\OIII]
emission along the spectral axis and $\pm$2.5 pixels in the spatial axis centered around
the expected location of the trace from the [\OIII] emission. This gives the 3$\sigma$ flux
limit of $6.4\times10^{-18}$ erg cm$^{-2}$ s$^{-1}$.

Table~\ref{tab4} lists the derived fluxes and corresponding luminosities of these lines.
 The values given are obtained without applying any dust correction or correction for slit losses. Both these will make the intrinsic luminosity of the [\OIII] line higher than what we infer. Till now [\OIII] emission is detected in 7 high-$z$ (i.e. \zabs$\ge$1.9) DLAs 
\citep[see Table~6 of][]{Fynbo13}. The observed [\OIII] luminosity in the present case
is close to the median value observed among other DLAs. The main difference is that the
present system has the highest \zabs\ and lowest metallicity among the high-$z$ DLAs with the detection of [\OIII]
emission.
The total [\OIII] luminosity is 0.3 L$_*$ as per the fit to the 
H$\beta$ +[\OIII] luminosity function given by \citet{Khostovan15} for $z\sim3$ galaxies.

\subsection{Emitting and absorbing gas kinematics}

\begin{figure}
  \centering
  \vbox{
  \includegraphics[width=1.0\linewidth]{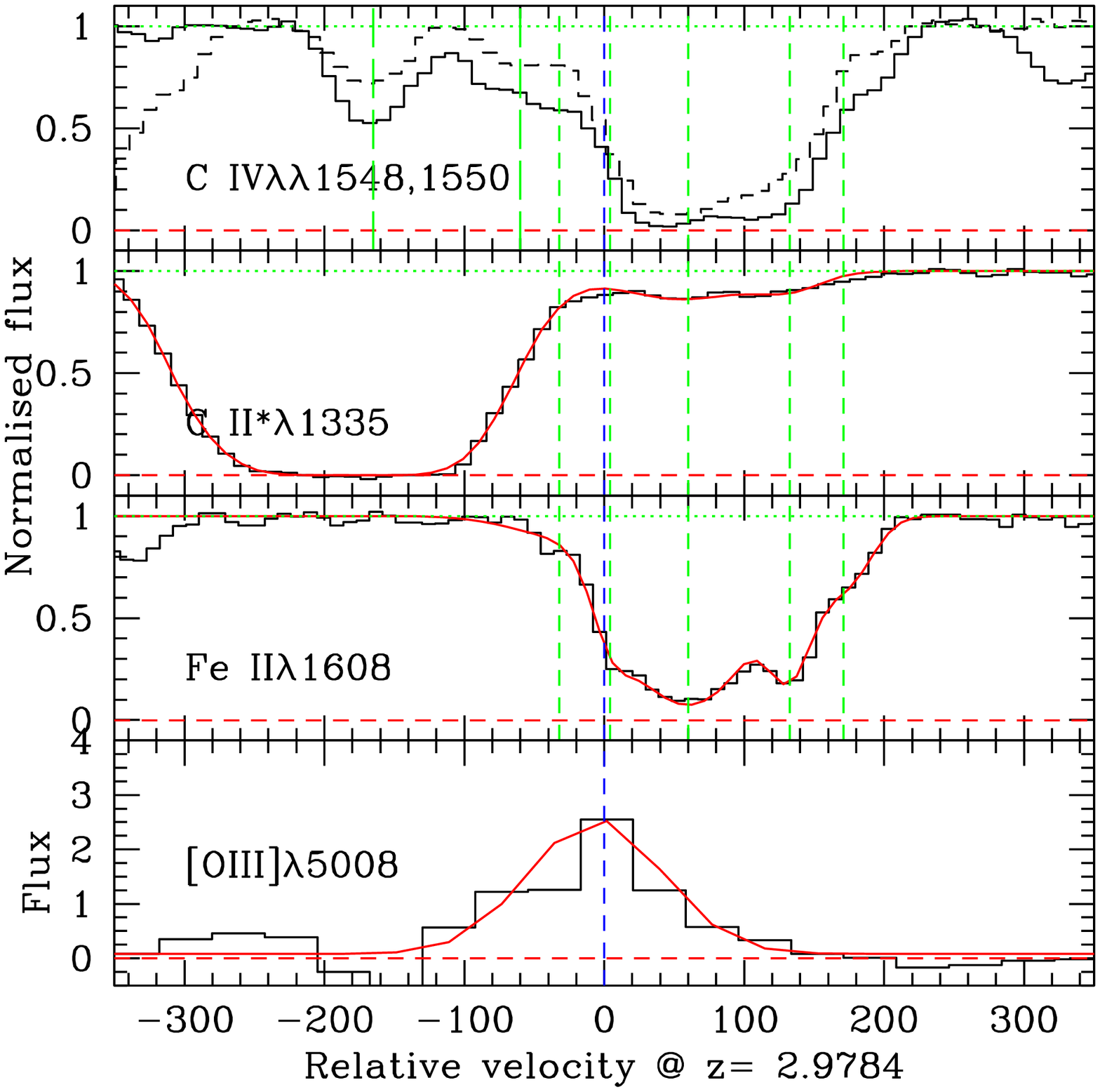}}
  \captionof{figure}{Comparison of absorption lines and [\OIII] \lam5008 emission associated with the
\zabs = 2.9791 DLA. The zero velocity is defined with respect to the  $z_{gal}$ = 2.9784 derived from
the [\OIII] emission. The bottom panel shows the [\OIII] emission (with flux shown in the unites of $10^{-17}$ 
ergs cm$^{-2}$ s$^{-1}$ \AA$^{-1}$) together with the best fitted Gaussian.
In the top panel, we plot  profiles of both the members of the \CIV\ doublet. Middle two panels
show profiles of \FeII\ (as a representative of the low ionization species) and \Cst\ together with 
the best fitted Voigt profiles (as also in Fig.~\ref{vpfit2}).  The vertical short-dashed lines show the 
individual components identified
for the low ions. Vertical long-dashed lines mark additional components seen only in \CIV\ absorption. }
  \label{ema}
\end{figure} 


The single component Gaussian fit to the [\OIII] lines gives a deconvolved FWHM of 110 \kms\ 
and a velocity dispersion of $\sigma\sim 46$ km s$^{-1}$. We can use this to estimate the
dynamical mass of the galaxy within the [\OIII] line emitting region. 
For a given optical size (${\rm r_{eff}}$) of 
the galaxy one can estimate the dynamical mass (${\rm M_{dyn}}$) using the 
relation ${\rm M_{dyn}\sim [3 \sigma^2 r_{eff}/G]}$ \citep[see][]{Maseda14}.
As we do not have the photometric measurement of ${\rm r_{eff}}$, we use two values
for estimating the range in the dynamical mass. First, we consider
${\rm r_{eff}}\le 4.5$ kpc as inferred
based on the fact that the emission is unresolved in the spatial direction along the slit.
This gives ${\rm M_{dyn}\le ~6.4 \times 10^9~M_\odot}$.
At $z\sim 3$, typical size of LBGs are smaller than 4.5~kpc limit we have found.
If we assume the typical ${\rm r_{eff}= 1}$ kpc \citep[as found by][for example]{Shibuya15}
measured for the LBGs at these redshifts we get  ${\rm M_{dyn}\sim 1.4 \times 10^9 ~M_\odot}$.
The inferred dynamical mass  is similar to what has been seen in the extreme emission line
galaxies at high-$z$ by \citet{Maseda13}.

Assuming a simple dark matter density profile $\rho(r)\propto r^{-2}$
(isothermal profile) and normalizing the density at the optical radius to the mean density 
within the optical radius (${\rm r_{eff}}$ = 1 or 4.5 kpc as considered above), we find the total mass 
within 12 kpc to be $3-4\times 10^{10} M_\odot$. 
This will mean a circular velocity of 105$-$121 \kms\ at 12 kpc. 
In Fig.~\ref{ema} we compare the emission line profile from the galaxy to the absorption 
profile measured along 
the QSO sight line. It is clear that almost all the low-ion absorption are redshifted with respect to the
systemic redshift of the galaxy defined by the centroid of the  [\OIII] emission
lines (i.e. \zem = 2.9784). Specifically, individual Voigt profile components used to fit the 
low ion absorption  have redshifted velocities of -32, 4.2, 60, 132.6 and 171.1 \kms.
In addition to these components \CIV\ absorption has components at $-$60 and $-$160 \kms.
Considerable absorption of low and high ions originate outside the circular velocity
we inferred above. Therefore, it is most unlikely that the absorption originates from 
an extended co-rotating disk.

It is possible that the absorbing gas may originate either from the
large scale infall or outflow as usually suggested for the {\it ``ultra-strong''} \MgII\ absorbers
\citep[][]{Gauthier12,Bouche12,Kacprzak12,Bouche13}. In these cases, galaxy rotational
velocity, orientation of the galaxy with respect to the QSO sight line are used in addition
to the absorption kinematics to draw conclusions on the nature of the gas flow.
In the absence of the galaxy image it will be difficult to draw firm conclusions on the nature 
of the gas flow in the present case.

In addition, the spatially unresolved [\OIII] emission we detect may originate from
a star forming clump that may be from a region away from the centre of a massive quiescent 
galaxy like the one found by \citet[][]{Zanella15}. Therefore, it is of utmost importance
to have deep imaging to get further insights into the nature of this galaxy.


\subsection{Star formation rate}

As we detect only [\OIII] lines we use the upper limits on [\OII] and H$\beta$ to infer
the upper limit on the star formation rate. We will use standard relationships used in
the literature assuming Salpeter IMF in this study. The calibration between [\OIII] luminosity
and SFR (derived from other tracers) is recently established in the case of high-$z$ LBGs
\citep[see for example,][]{Suzuki15} and GRB host galaxies \citep{Kruhler15}. If we use these 
relationships we get 6.6$\le$SFR~(M$_\odot$ yr$^{-1}$)$\le 25$. 
From the 3$\sigma$ limit on the [\OII] luminosity we derive a star formation rate (SFR) of
$\le 2.1$ M$_\odot$ yr$^{-1}$  using the relationship given by \citet{Kennicutt98}. Similarly if
we use the standard Balmer ratio of 2.8 and the relationship between H$\alpha$ luminosity
and SFR given by \citet{Kennicutt98} we derive SFR $\le$ 3.8 M$_\odot$ yr$^{-1}$ for the
3$\sigma$ upper limit on the H$\beta$ luminosity. The wide range in SFR derived above
reflects the fact that ionized gas in the present case is highly excited compared to what is typically
seen. We discuss this in detail in the next section.

We can also estimate the upper limit on SFR in the DLA galaxy using the inferred upper 
limit on the \lya\ line flux assuming that the \lya\
photons mainly originate from the \HII\ regions around massive stars
and case B recombination \citep{Osterbrock06}. 
The \lya\ luminosity (L$_{Ly\alpha}$) is then related to the SFR ($\dot{M}_{SF}$) by,
\begin{equation}
L(Ly\alpha) = 0.68~h\nu_\alpha~(1-f_{\rm esc})~N_\gamma~\dot{M}_{SF}
\end{equation}
where $h\nu_\alpha$ = 10.2 eV, $f_{\rm esc}$ and $N_\gamma$ 
are, respectively, the energy of a \lya\
photon, the escape fraction of Lyman continuum photons
and the number of ionizing photons released per baryon
of star formation. We
use  $f_{\rm esc}=0.1$ and $N_\gamma = 7880$ that is appropriate
for the measured metallicity of the DLA \citep[from Table 1 of][]{samui07}.
Thus the observed \lya\ luminosity gives an upper limit on the star formation rate,
${\dot{M}_{\rm SF}\le 4~ (f_{\rm esc}^{\rm Ly\alpha}/0.05) M_\odot yr^{-1}}$. 

The non-detection of stellar continuum in the visible part of the spectrum places
a constraint on the flux at the rest frame 1500~\AA\ of the galaxy. However the implied limit
on the star formation rate is not stringent (i.e. $\le 27~M_\odot yr^{-1}$).  All this suggests that the
  DLA galaxy is forming stars at moderate rate. The strong [\OIII] emission seen in this
  system may be related to the high excitation in the ionized gas. We discuss this in detail in the
  following section.

\subsection{High [\OIII]/[\OII] ratio and metallicity}

The most interesting aspect of the present system is the very large value (i.e. $\ge10$) 
of the [\OIII]/[\OII] and [\OIII]/[H$\beta$] ratios. It is well documented now that $z>2$ galaxies tend to have
elevated [\OIII]/[\OII] ratio compared to local star forming galaxies \citep[see for example,][]{Steidel14,Masters14}. However, galaxies with this ratio greater than 10 are rare. 

In the local universe, such high ratios (in the range 10-50) are seen in `extreme blue compact dwarf galaxies (BCDs)'. It has also been pointed out that these extreme BCDs are compact,  have low metallicity and 
high specific star formation as found in high-$z$ LBGs. High excitations seen in low-$z$ BCDs
and high-$z$ galaxies may be attributed to many possibilities such as low metallicity, high ionization 
parameter, hard ionizing radiation field and/or the presence of a density-bound 
\HII\ regions \citep[][]{Stasinska15}.

Narrow velocity width of the [\OIII] line probably rules out the possibility of excitation due to a hidden AGN.
Also the absence of [\OII] line rules out the strong [\OIII]/[H$\beta$] being due to elevated oxygen
abundance. Recently it was suggested that systems with elevated [\OIII]/[\OII] ratio may have large
Lyman continuum (LyC) escape fraction if the gas is optically thin (i.e. matter bound \HII\ regions). Such
galaxies will also show strong \lya\ emission as it's escape fraction is also enhanced due to optical depth being low
\citep[see for example,][]{Jaskot14,DeBarros15}. Indeed, in the \zem = 3.2 Lyman continuum leaking galaxy
studied by \citet{DeBarros15} the observed $L[\OIII]/L[\OII]$ and $L[\OIII]/L[H\beta]$ ratios are consistent
with our galaxy. However, \lya\ emission is clearly detected with $L[\OIII]/L(Ly~\alpha) = 1.4$. Our observations
rule out such a ratio by more than 7$\sigma$ level. Therefore, it is most unlikely that the
elevated [\OIII] luminosities seen in the present case may be due to matter bound line emitting nebula.

\citet{Stanway14} have argued that the high [\OIII]/H$\beta$ ratio can be explained through the ageing
of a rapidly formed stellar population. The probability of detection of galaxies with high [\OIII]/H$\beta$
depends on the time-scale over which such elevated ratios are maintained in the star bursting region. They 
showed that the inclusion of binary evolution in the stellar synthesis code enhances this duration up to few
100 Myrs. It is interesting that even in their model one needs high densities and low metallicity 
in the ISM to get such large [\OIII]/H$\beta$ ratios. In this scenario 
the present system could 
have gone through a recent star burst activity with a relatively metal poor 
ageing stellar population surrounded by a dense interstellar medium.

Using the {\sc IZI} (Inferring the gas phase metallicity (Z) and ionization parameter of the
ionized nebula) code described in \citet{Blanc15} and assuming the photoionization model results
of \citet{Levesque10} we derive constraints on the nebular metallicity (i.e. 12+[O/H] $\le$ 8.5),
ionization parameter (i.e. log~q$>$8.14 and log~U$\ge-2.33$). Such large ionization parameters
are inferred in $\le$ 25\% of the $z\ge2$ galaxies studied by \citet{Masters14}. The upper limit on
metallicity is consistent with the metallicity we derive for the DLA along the QSO
line of sight. While we need to detect other nebular lines with deep spectroscopic observations
to draw firm conclusions on the physical conditions in this DLA galaxy, all indications are
suggesting the galaxy to be compact, moderately star forming and having low metallicity
and high specific star formation rate.

\subsection{\Cst\ cooling rate:}

Associated \Cst\ \lam1335 absorption is clearly detected in this DLA (see Fig.~\ref{ema}). 
Using Voigt profile fits we obtain $\log N$(\Cst) = 13.64$\pm$0.07. This 
together with the total $N$(\HI) measured gives a cooling rate $\log~l_c=-27.57\pm0.12$
ergs s$^{-1}$ H$^{-1}$. \citet{Wolfe08} have proposed a bimodal distribution in 
DLA population  based on [\CII] cooling rate $l_{c}$ : ``low-cool'' DLAs with $l_{\rm c} \leq l_{\rm c}^{\rm crit}$ and 
``high-cool'' DLAs with $l_{\rm c} > l_{\rm c}^{\rm crit}$ (where $l_{\rm c}^{\rm crit} = 10^{-27} \rm ergs ~\rm s^{-1} \rm H^{-1}$). The cooling rate inferred for the present DLA belongs to the ``low-cool'' population.
The metallicity and cooling rate inferred in the present case is close to that of the \zabs = 2.5397 system
toward J1004+0018 studied by \citet{Dutta14}. We can use the model results presented in their Fig.~14
to interpret \Cst\ observations for the present system. The observed ratio,
$\log N$(\Cst)/$N$(\SiII) = $-1.85\pm0.25$ and the cooling rate can be explained with a radiation field
similar to Galactic mean UV field and a hydrogen density of 1 cm$^{-3}$. Therefore, the absorbing
gas (if part of a cold neutral medium) is seeing excess radiation either from the galaxy 
or from the ongoing in situ star formation.

Following \citet{Wolfe08} we can write,
\begin{equation}
l_c = 10^{-5} \kappa \epsilon J_\nu,
\end{equation}
where, $\kappa = 0.018$, $\epsilon$ and $J_\nu$ are dust to gas ratio, grain heating efficiency and local background radiation field intensity respectively. When we use the maximum dust efficiency (i.e. 2\%) 
in the cold neutral medium found by \citet{Weingartner01} we get $J_\nu^{\rm local} = 3.1 \times 10^{-20}$ erg s$^{-2}$ cm$^{-2}$ Hz$^{-1}$.  
If we further use equation 3 of \citet{Wolfe08} we get the in-situ surface star formation rate,
$\Sigma_{SFR}=2\times10^{-4}$ M$_\odot$ yr$^{-1}$ kpc$^{-2}$. Although very small, our present observations can not
rule out such a low SFR along the QSO sightline. Alternatively if the gas is not part
of the stellar disk (as it is most likely) then the photo-heating can come 
from the galaxy light and in particular the
[\OIII] emitting region. Such a scenario has been proposed to explain the physical
conditions in low-$z$ quasar-galaxy pairs \citep[see for example,][] {Dutta15}.

However, if absorbing gas detected along the QSO sight line is part of an infalling or 
outflowing gas, then \Cst\ may be originating
from a warm or partially ionized gas as suggested by \citet{Srianand05} in the case of 
H$_2$ bearing DLAs. 
If that is the case then we will not have much handle on the star 
formation rate from the \Cst\ cooling rate. This again suggests that purely
based on the \Cst\ detection alone we will not be able to conclude that the absorbing
gas is part of the cold ISM.

\section{\lya\ emission from the DLA at \zabs = 3.2477}\label{dgal2}

\begin{figure*}
 \centering
  \includegraphics[bb = 150 10 450 800,height=18cm,angle=90,clip=true]{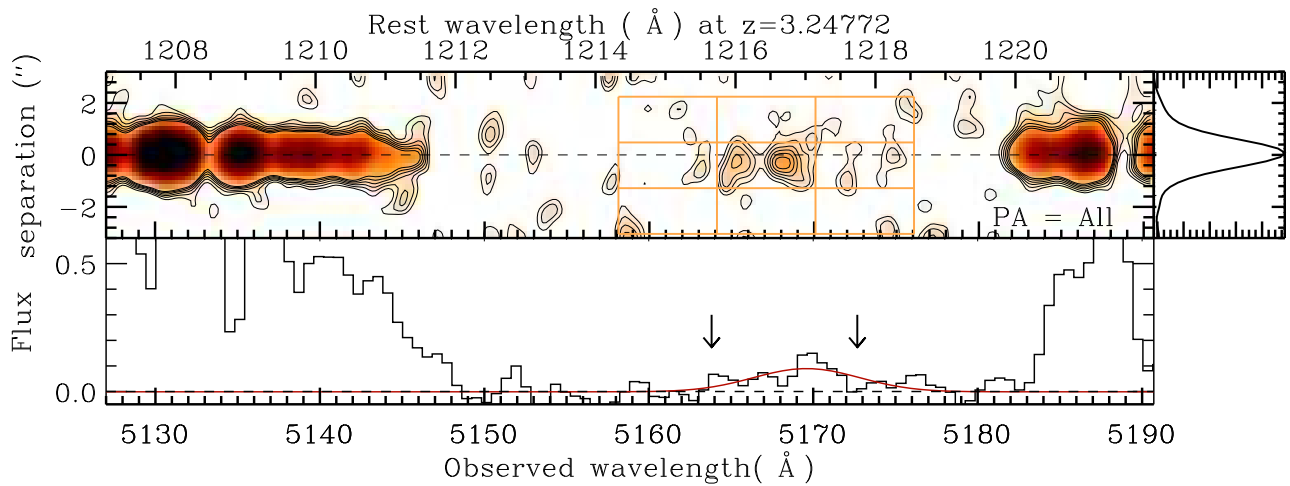} 
 \caption{The \lya\ emission from \zabs = 3.2477 DLA towards J2358+0149. 
In the top panel we show the 2D image of the QSO in the wavelength range of the \lya\ 
emission observed in the combined 2D spectra obtained with three slit
orientations.  We clearly see excess flux in the regions identified by the 
middle box. In the bottom panel, we plot the extracted 1D spectrum and a 
single component Gaussian fits to the residual flux in the bottom of the DLA profile. 
The two arrows show the locations of the metal absorption redshift
and QSO emission redshift.}             
 \label{lyaem}      
 \end{figure*} 

In the combined 1D spectrum we see some residual flux at the bottom of the 
\lya\ line of \zabs =  3.2477 DLA (see bottom panel in Fig.~\ref{lyaem}). 
In the top panel of Fig~\ref{lyaem} we plot the 2D-Gaussian smoothed (with a
FWHM of 5 pixels along the spatial direction and 3 pixels along the observed wavelength) 
combined 2D spectrum obtained with slits aligned at three 
different position angles. The overlayed contours are at the flux levels 
2, 4, 6, 8, 10, 12, 14 $\times 10^{-20}$ ${\rm erg~s^{-1}~cm^{-2}}$~\AA$^{-1}$. 
Excess flux is clearly seen in the wavelength range where we found
residuals in the 1D spectrum.
It is also clear from the images that most of the residual flux found is close to the
trace suggesting that the \lya\ emitting source is very close to our line of sight
to the QSO. In order to estimate the significance of the \lya\ emission
we considered 9 apertures as shown by boxes in Fig.~\ref{lyaem}.
Each aperture is 6 \AA\ wide in the observed wavelength and
1.75 arc sec in the spatial direction.  As expected the integrated flux
(${\rm 4.5\times10^{18} erg~s^{-1}~cm^{-2}}$) is maximum for the middle aperture
and is 6 times higher than the average flux [i.e. ${\rm (0.72\pm0.17)\times10^{-18} erg~s^{-1} ~cm^{-2}}$] 
of the remaining 8 apertures. As can be seen from
the figure the peak emission is $\sim 0.4$ arc sec away from the center
of the QSO trace. 

In order to understand the possible spatial separation between the
\lya\ emission and the QSO we repeated the above exercise on the 2D
spectra obtained at different slit orientations. In the case of
spectra obtained with position angles (PA) 0$^\circ$ and -60$^\circ$ we detect
maximum flux in the middle aperture with the mean flux of the 
other 8 apertures less by about a factor of 3.  We do not detect significant 
excess emission in the middle
aperture in the case of PA = 60$^\circ$. But the flux errors are large
and flux measured in other two position angles are consistent within
1.5$\sigma$ for the non-detection.
In the case of
PA = 0$^\circ$ the peak emission is at 0.5 arc sec below the QSO
trace. In the case of PA = -60$^\circ$ the emission is concentrated
in two blobs one above and one below the trace (with the off-set of
$\le$ 0.6 arc sec). This suggests that the
\lya\ emission is extended and off-centred with respect to the QSO
position. While we need a better S/N spectrum to perform triangulation
the present data suggests that the emitting region is within $\sim$5 kpc
of the QSO line of sight.

It is well known that, due to complex radiative transport, the profile of the
\lya\ emission will not be a simple Gaussian. However, for simplicity 
we fit the \lya\ line in the 1D spectrum with a single Gaussian. The
fit is also shown in the bottom panel of Fig.~\ref{lyaem}. The
centroid of the Gaussian gives \zem = 3.2512$\pm$0.0004 and the integrated
\lya\ flux is (5.04$\pm$0.90)${\rm \times ~10^{-18} ~ergs~cm^{-2}~s^{-1}}$.
This confirms the emission at the 5$\sigma$ level. This corresponds to a 
luminosity of (4.5$\pm$0.8)${\rm \times 10^{41}~erg~s^{-1}}$.
Instead of fitting the Gaussian, if we simply measure the total flux
in the bottom of the \lya\ absorption trough, we find the 
total flux to be (7.2$\pm$1.3)${\rm \times10^{-18}  ~ergs~cm^{-2}~s^{-1}}$
and the corresponding luminosity is $(6.4\pm1.2){\rm \times 10^{41}~erg~s^{-1}}$.
\citet{Finley13} have found a strong \lya\ emission in 25\% of the pDLAs with
$\log N$(\HI)$\ge$21.3. Their analysis favor these DLAs being associated with
the host galaxies of the QSOs and probably not completely covering the \lya\
from the narrow line regions (NLR). The \lya\ flux measured in the present case
is nearly 35 times less than the average flux measured by \citet{Finley13} and
probably do not belong to the pDLA population identified by them.

\begin{figure}
  \centering
  \vbox{
  \includegraphics[width=1.0\linewidth]{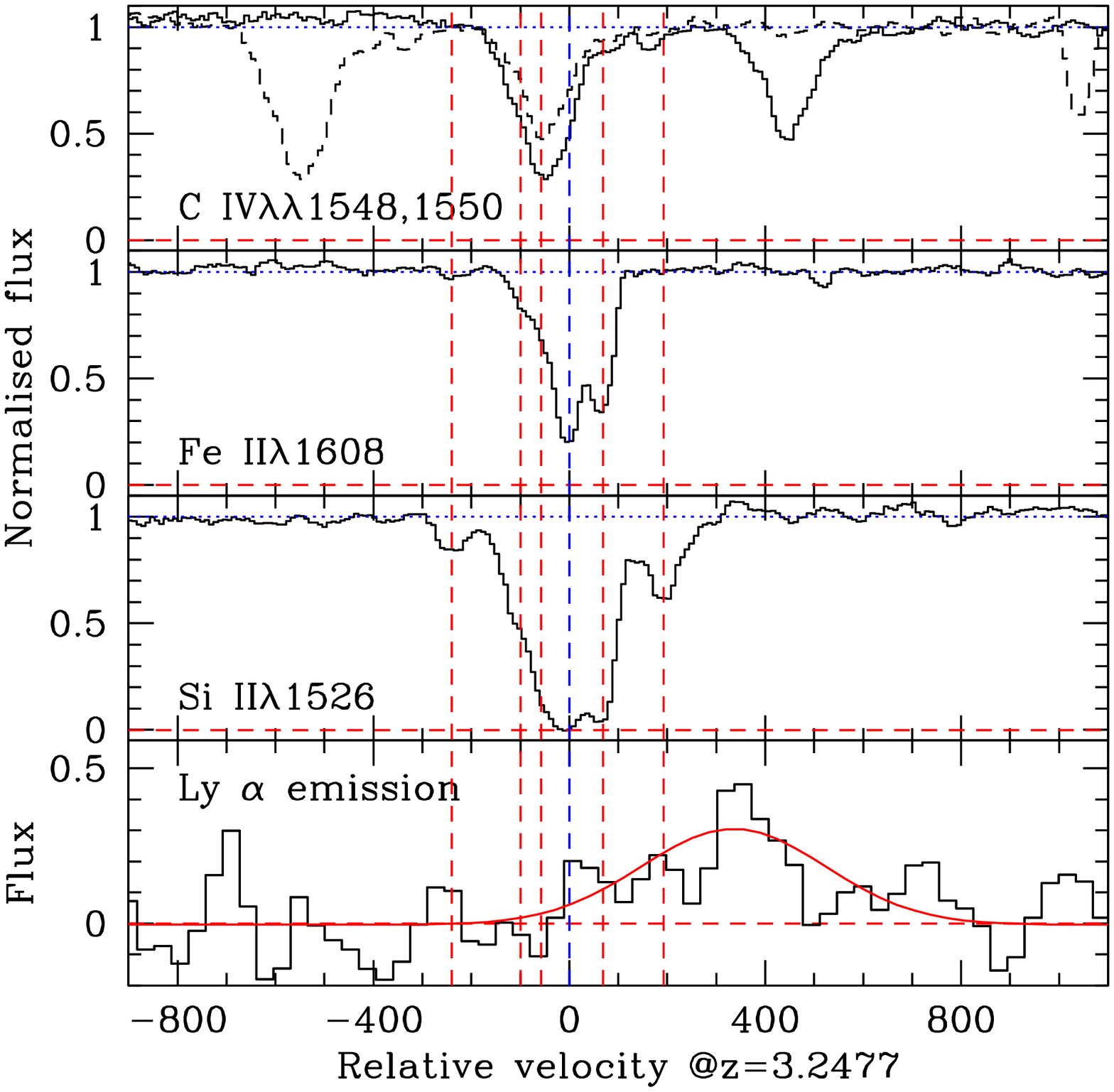}}
  \captionof{figure}{
Comparison of absorption lines and the \lya\ emission associated with the
\zabs = 3.2377 DLA. The zero velocity is defined with respect to the  $z_{abs}$ = 3.2477. The bottom panel shows 
the \lya\ emission (with flux shown in the unites of $10^{-17}$ 
ergs cm$^{-2}$ s$^{-1}$ \AA$^{-1}$) together with the best fitted Gaussian.
In the top panel, we plot  profiles of both the members of the \CIV\ doublet. Middle two panels
show profiles of \FeII\ and \SiII\ (as a representative of the low ionization species).  
The vertical short-dashed lines show the individual components identified
for the low ions.} 
  \label{ema1}
\end{figure}

The \lya\ emission is roughly 280 \kms\ blue shifted with respect to the
QSO based on the redshift of the QSO measured from the Mg~{\sc ii} 
emission line peak. However, the \lya\ emission is redshifted by about 
320 \kms\ with respect to the strongest absorption component.  For the 
observed value of $N$(\HI) for this system, if we assume the
DLA to be a static medium having a kinetic temperature of 10$^4$ K 
then we expect the \lya\ emission to have a
double hump with the peak in the red wing shifted by about 375 \kms\
\citep[see][]{Dijkstra14}. This is roughly consistent with the 
shift we notice above between the redshift of the \lya\ emission 
and metal absorption lines.  It is also interesting to note that
the peak \lya\ emission occurs just outside the maximum velocity of
the low ions (see Fig.~\ref{ema1}) consistent with what one would
have expected based on the \lya\ line transport.

In a simple radiative transport model, this can be understood in terms of scattering from an expanding (i.e. outflowing) \HI\ shell.
However the actual radiative transport
may be a bit more complex when the medium is not static and has 
dust.
In addition, the \lya\ emission could come either
from the star forming regions in the DLA or from the fluorescence 
induced by the QSO UV flux if the absorber is very close to the QSO.
We discuss both the possibilities.

\subsection{Star formation rate}

First we consider the possibility that the observed \lya\ emission
originates from an intervening galaxy. In that case,
we estimate the average SFR in the DLA, ${\rm \dot{M}_{SF}\sim0.17-0.23~M_\odot yr^{-1}}$, 
using the prescription described in Section~4.2. 
The range comes from
the luminosities estimated using the simple integration and the single component
Gaussian fit to the \lya\ emission.
Only a small fraction of 
\lya\ generated in the galaxy escapes the galaxy and some fraction of this
emerging \lya\ photons are also absorbed by the IGM.  Therefore, the above
derived star formation rate without making any corrections for the radiative
transport should be considered as a lower limit. 

We detect neither [\OII] nor [\OIII] emission. As the expected position
of H$\beta$ coincides with the H$\beta$ emission from the QSO it is
difficult to estimate the H$\beta$ luminosity associated to the DLA.
From the rms in the expected position of the [\OII] line in the QSO
continuum we obtain a 3$\sigma$ upper limit on the [\OII] luminosity of
1.2$\times 10^{41} (3\sigma)$ erg s$^{-1}$. If we use the relationship
given by \citet{Kennicutt98} this translates to an upper limit in the
star formation rate of 1.7 ${\rm M_\odot yr^{-1}}$. This implies that
the \lya\ escape fraction is more than 10\% if we use the \lya\
luminosity based on Gaussian fits or $\ge 14$\% if we use the 
integrated \lya\ luminosity. Inclusion of corrections for dust extinction 
and the \lya\ opacity of the IGM will further increase this lower limit.

The lower limit we find here is higher than the mean measured value 
(i.e. $f_{\rm esc}^{\rm Ly\alpha}\sim 5\%$) in the high-$z$ LBGs \citep[][]{Hayes10}.
However, such high values of $f_{\rm esc}^{\rm Ly\alpha}$ are also seen in a couple 
of high-$z$ DLAs [i.e.  $f_{\rm esc}^{\rm Ly\alpha}= 55\%$ in the case of 
\zabs = 2.35 DLA towards Q~2222-0946 \citep{Fynbo10} and 
$f_{\rm esc}^{\rm Ly\alpha}=20\%$ in the case of \zabs = 2.21 DLA towards Q1135-0010
\citep{Noterdaeme12}]. However, unlike the present case the inferred SFR
based on the emission lines detected in the NIR in these cases are very high
(i.e. $\ge 10$ M$_\odot$ yr$^{-1}$).

\subsection{Fluorescent \lya\ emission induced by the quasar}

The fluorescent \lya\ emission induced by the QSO is a viable alternative, in particular if
the escape fraction of \lya\ is very small. We explore this possibility here.

To start with, we note that the size and surface brightness of the \lya\ emission
seen are both less than what is typically seen in the case of radio-loud QSOs at
similar redshifts \citep[see for example,][]{Roche14}.This suggests that the origin of 
\lya\ emission in the present case may be very different from the extended \lya\ 
seen around radio galaxies.  We also note that,
based on the surface brightness of the \lya\ emission, lack of \NV\ absorption
and fine-structure lines of \SiII, this DLA is not similar to the one 
studied by \citet[][]{Fathivavsari15}, where the DLA is thought to be dense,
compact and close to the QSO covering the broad as well as narrow emission line
region only partially \citep[see also,][]{Finley13}. Therefore, it is most likely that
we are not seeing the extended NLR of the host galaxy. 

For the set of  cosmological parameters assumed in this work, if we consider
the redshift difference between the DLA and the QSO to be due to 
spatial separation then we expect the DLA to be 1.8 Mpc away from the QSO.
As some part of the redshift difference may come from the peculiar 
velocities the actual separation can be slightly different.

From our flux calibrated spectrum and SDSS spectrum we infer the flux at 
912~\AA\ in the rest frame of the QSO to be 2.3$\times 10^{-16}$ erg cm$^{-2}$ s$^{-1}$ \AA$^{-1}$.
Assuming the UV spectrum of the QSO to be a power-law, $J_\nu \propto \nu^{-1.4}$, we estimate
the \HI\ photoionization rate as a function of distance  in Mpc, $r_{\rm Mpc}$, as
$\Gamma_{\rm HI}^Q = 1.14\times10^{-11}/r_{\rm Mpc}^2 (s^{-1})$. From the recent computations of
UV-background using updated QSO and galaxy emissivities \citep[see][]{Khaire15,Khaire15a},
 we estimate
the hydrogen photoionization rate due to the background to be 
$\Gamma_{\rm HI}^{\rm bgr} = 6.9\times10^{-13} (s^{-1})$. The two rates are equal for $r_{\rm Mpc} = 4.06$. 
Therefore, in the absence of in situ star formation,
the DLA will receive at least 5 times more ionizing photons from the QSO compared to that from the
background. Therefore, at face value the observed \lya\ originating from fluorescence is
a realistic possibility.

The fluorescence induced \lya\ emission around radio-quiet QSOs have been reported in a few cases
\citep[see for example,][]{Adelberger06, Francis06,Cantalupo12, Cantalupo14}. Typically
one looks for: (i) large \lya\ line equivalent width mainly due to the absence of a strong continuum,
(ii) profiles showing double humps and (iii) large surface brightness, to 
identify the candidate fluorescence \lya\ emitters. As the \lya\ emission is very close
to the QSO trace, it is very difficult to measure the equivalent width in the present case.
We do observe that the \lya\ emission peak is shifted with respect to the metal line absorption
from the DLA. However this alone will not confirm the fluorescence as such a profile is
also expected from the radiative transport even in the case of \lya\ induced by  the
embedded stars.

Following \citet{Shull14}, we can write the integrated unidirectional flux of the 
\HI ionizing photons, $\phi_0 = \int_{\nu_0}^{\infty} \pi I_\nu d\nu/h\nu$.
We estimate $\phi_0 \sim 84960$ photons for the meta-galactic UV background at $z\sim3.2$. 
If we assume a one arcsec$^2$ optically thick slab of \HI\ gas in photoionization
equilibrium with this background 
then we get a surface brightness, $SB_{\rm Bg} \sim 1.9\times 10^{-20}$ erg s$^{-1}$ cm$^{-2}$ arcsec$^{-2}$,
if the disk is seen face-on. Otherwise there will be a dilution factor that is a ratio of projected 
area along the line of sight to the actual area.
Here, $SB_{\rm Bg}$ is the surface brightness expected purely
due to the UV background radiation.

The observed surface brightness of the \lya\ emission in the present case is 2.9$\times10^{-18}$
erg s$^{-1}$ cm$^{-2}$ arcsec$^{-2}$. 
 In general,  the surface brightness induced by the
QSO fluorescence can be written as $SB = (1+b)~SB_{\rm Bg}$, where the factor $b$ is 
just $\Gamma_{\rm HI}^Q/\Gamma_{\rm HI}^{\rm Bg}$.  In the present case, for the face on condition,
we find $b\sim$142 and the cloud has to be at a distance of 340 kpc from the QSO.
Therefore, the \lya\ emission seen in the present case can be produced by
the QSO fluorescence if the gas is at $\sim$340 kpc from the QSO. 
However, the exact
value will depend on the assumed geometry of the absorbing gas and the projected area
towards the QSO and along the line-of-sight.
Spatially resolved detection of associated stellar light,
better mapping of the \lya\ emission and detection of other emission lines with deeper
spectroscopic observations will allow us to test the fluorescence scenario more thoroughly.


\section{Summary and discussions}\label{conc}

Using long-slit spectroscopic observation of J2358+0149 (\zem = 3.255), obtained in three different position angle with
VLT-X-shooter, we have searched for emission lines originating from four DLAs and one sub-DLA in the redshift 
range 2.73$-$3.25. In this work we have presented the column density, metallicity and depletion measurements 
for all the 5 systems.
We reported the detection of emission lines associated with two DLAs having, $\log N(\HI) > 21.0$, with low ion 
absorption lines showing large velocity widths (i.e. $\Delta v_{90}>140$ \kms for the \FeII\ \lam1608 line). We do not detect any emission in the remaining three systems that have low $N$(\HI) and narrow metal line widths.

In the case of \zabs = 2.9791 ESDLA, that also satisfies the definition of ``{\it ultra-strong}'' \MgII-systems, we detect
[\OIII] \lam\lam4960,5008 emission at a projected separation of 11.9$\pm$0.8 kpc from the QSO sight line. The absorbing gas
has a metallicity of [Zn/H] = $-1.83\pm0.18$ and moderate dust depletion (i.e. [Zn/Fe] = $0.15\pm0.25$).
The absence of 
[\OII] and H$\beta$ emission suggests that the galaxy is a high-excitation galaxy, similar to the ``{\it extreme blue compact
dwarf}'' galaxies seen at the low-$z$, undergoing a moderate star formation (SFR ${\rm \le 2.1~{M_\odot} yr^{-1}}$). 
The large, [\OIII]/[\OII]$\ge$10 and [\OIII]/H$\beta$ $\ge$10 ratios seen, are very 
rare even among high-$z$ LBGs that usually show elevated ratios compared to their low-$z$ counterparts. Based on
the absence of \HeII, \CIV, \lya\ and \MgII\ emission we feel the hidden AGN contribution to the excitation is least
likely. We also do not favor the matter bound ionized region scenario as one would expect a strong \lya\ emission in this 
case. Considerable progress can be made if one can detect other emission lines and image 
the galaxy in the continuum light using deep observations. This will allow us to measure the metallicity and ionization 
parameter of the nebula and constrain the nature of the {star formation in this galaxy}. In addition, one will be able to understand the origin of the 
large velocity spread seen in the absorption lines in terms of cold accretion or large scale outflows. This will allow
us to build a consistent model of this system.

In the case of \zabs = 3.2477 proximate DLA we detect extended and diffuse \lya\ emission in the DLA trough. The DLA has
a metallicity of [Zn/H] = $-0.97\pm0.13$ and moderate depletion [Zn/Fe] =$0.53\pm0.20$. As in the previous case the 
metal line absorption has a very large spread and observed \MgII\ equivalent widths are consistent with the 
system being called an ``{\it ultra-strong}'' \MgII-system. The peak of the \lya\ emission is redshifted by about 330 \kms\
with respect to the strongest low-ion absorption component. This is consistent with what is expected from the simple radiative
transport models with gas outflow. As the absorber is very close to the systemic redshift of the QSO we have two viable scenarios for the origin of \lya\ emission: from in situ star formation in the DLA galaxy or \lya\ fluorescence induced by the QSO.
In the former case we use the \lya\ luminosity to estimate the star formation rate in the range 0.17$-$0.23 M$_\odot$ yr$^{-1}$ assuming 
$f^{\rm Ly~\alpha}_{\rm esc} = 1$. Based on the lack of [\OII] emission line we derive SFR$\le1.7$ M$_\odot$ yr$^{-1}$ and
$f^{\rm Ly~\alpha}_{\rm esc} \ge 0.29$. The latter is higher than what is typically measured in high-$z$ LBGs \citep[][]{Hayes10} but
consistent with what is seen in the case of \zabs=2.207 DLA towards SDSS~J113520.39-001053.56  \citep[][]{Noterdaeme12}. 
We show that the \lya\ fluorescence, caused by the QSO, can also reproduce the observed \lya\ emission, provided the absorbing gas lies no more than 340 kpc from the QSO.
While the exact physical situation in the present case may not be similar to what is seen in the radio-loud QSOs or in
the compact proximate DLAs that do not cover the narrow line emission regions, present data does not rule out the
fluorescence scenario. Detecting or placing a deep limit on the continuum emission is important to choose between the
two viable alternatives discussed here. High resolution spectra of the QSO is also needed to get accurate
  metallicity and gas kinematics based on the absorption lines.

\section{Acknowledgment}
TH wishes to thank IUCAA for hospitality. We thank Sussana Vergani for her help in preparing
the observational blocks.{We thank the anonymous referee for useful comments.}

\def\aj{AJ}%
\def\actaa{Acta Astron.}%
\def\araa{ARA\&A}%
\def\apj{ApJ}%
\def\apjl{ApJ}%
\def\apjs{ApJS}%
\def\ao{Appl.~Opt.}%
\def\apss{Ap\&SS}%
\def\aap{A\&A}%
\def\aapr{A\&A~Rev.}%
\def\aaps{A\&AS}%
\def\azh{AZh}%
\def\baas{BAAS}%
\def\bac{Bull. astr. Inst. Czechosl.}%
\def\caa{Chinese Astron. Astrophys.}%
\def\cjaa{Chinese J. Astron. Astrophys.}%
\def\icarus{Icarus}%
\def\jcap{J. Cosmology Astropart. Phys.}%
\def\jrasc{JRASC}%
\def\mnras{MNRAS}%
\def\memras{MmRAS}%
\def\na{New A}%
\def\nar{New A Rev.}%
\def\pasa{PASA}%
\def\pra{Phys.~Rev.~A}%
\def\prb{Phys.~Rev.~B}%
\def\prc{Phys.~Rev.~C}%
\def\prd{Phys.~Rev.~D}%
\def\pre{Phys.~Rev.~E}%
\def\prl{Phys.~Rev.~Lett.}%
\def\pasp{PASP}%
\def\pasj{PASJ}%
\def\qjras{QJRAS}%
\def\rmxaa{Rev. Mexicana Astron. Astrofis.}%
\def\skytel{S\&T}%
\def\solphys{Sol.~Phys.}%
\def\sovast{Soviet~Ast.}%
\def\ssr{Space~Sci.~Rev.}%
\def\zap{ZAp}%
\def\nat{Nature}%
\def\iaucirc{IAU~Circ.}%
\def\aplett{Astrophys.~Lett.}%
\def\apspr{Astrophys.~Space~Phys.~Res.}%
\def\bain{Bull.~Astron.~Inst.~Netherlands}%
\def\fcp{Fund.~Cosmic~Phys.}%
\def\gca{Geochim.~Cosmochim.~Acta}%
\def\grl{Geophys.~Res.~Lett.}%
\def\jcp{J.~Chem.~Phys.}%
\def\jgr{J.~Geophys.~Res.}%
\def\jqsrt{J.~Quant.~Spec.~Radiat.~Transf.}%
\def\memsai{Mem.~Soc.~Astron.~Italiana}%
\def\nphysa{Nucl.~Phys.~A}%
\def\physrep{Phys.~Rep.}%
\def\physscr{Phys.~Scr}%
\def\planss{Planet.~Space~Sci.}%
\def\procspie{Proc.~SPIE}%
\let\astap=\aap
\let\apjlett=\apjl
\let\apjsupp=\apjs
\let\applopt=\ao
\bibliographystyle{mnras}
\bibliography{thbib}

\appendix
\label{lastpage}
\section{Voigt profile fits to the metal lines:}

\begin{figure}
  \centering
  \vbox{
 \includegraphics[bb=30 15 550 785,width=1.0\linewidth,clip=true]{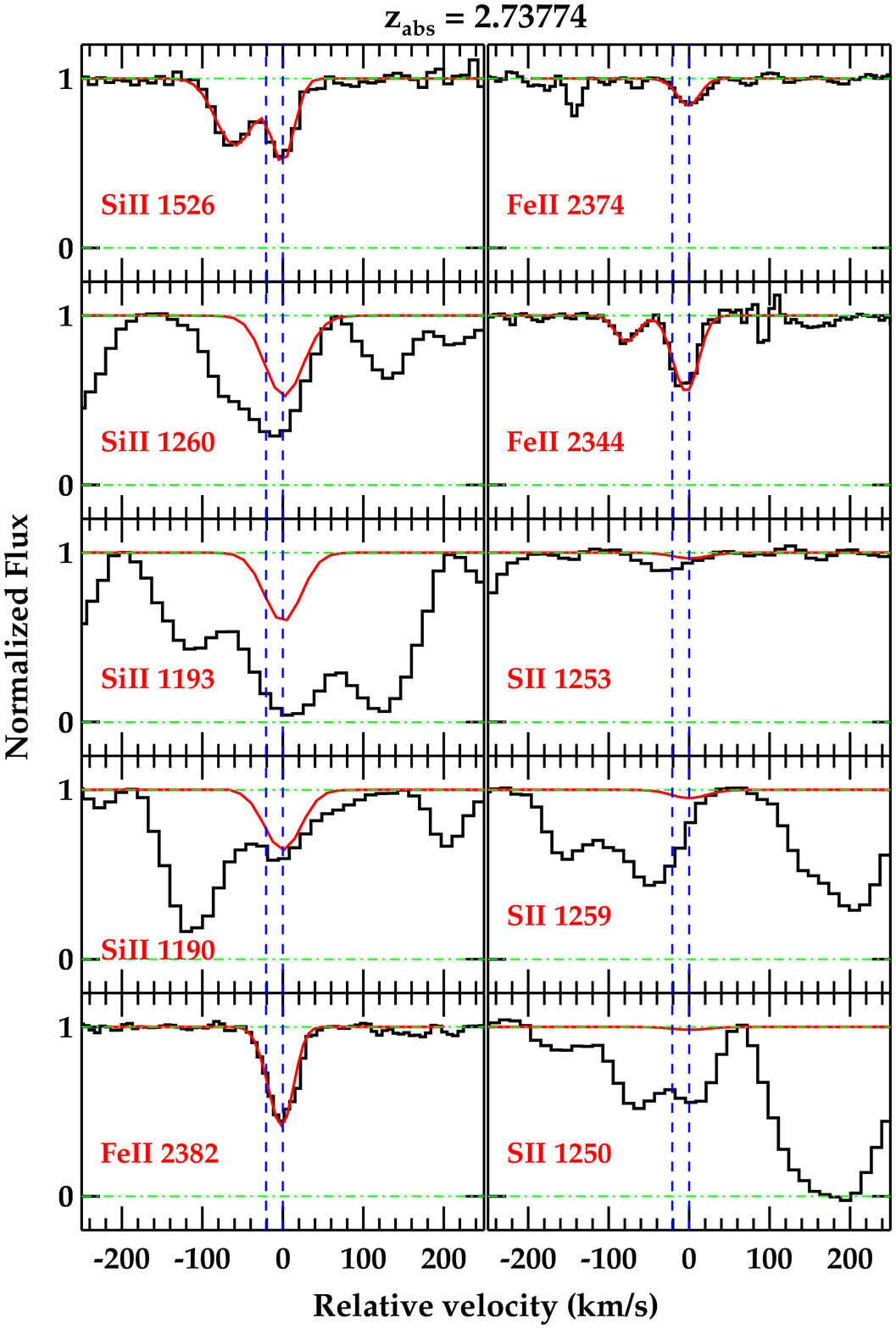}
 }

 \caption{Velocity plot of low ion absorption lines detected in \zabs=2.7377 sub-DLA together with the best fitted Voigt profiles.
}
 \label{vpfit1}
\end{figure} 

\begin{figure}
  \centering
  \vbox{
 \includegraphics[bb=30 15 550 785,width=1.0\linewidth,clip=false]{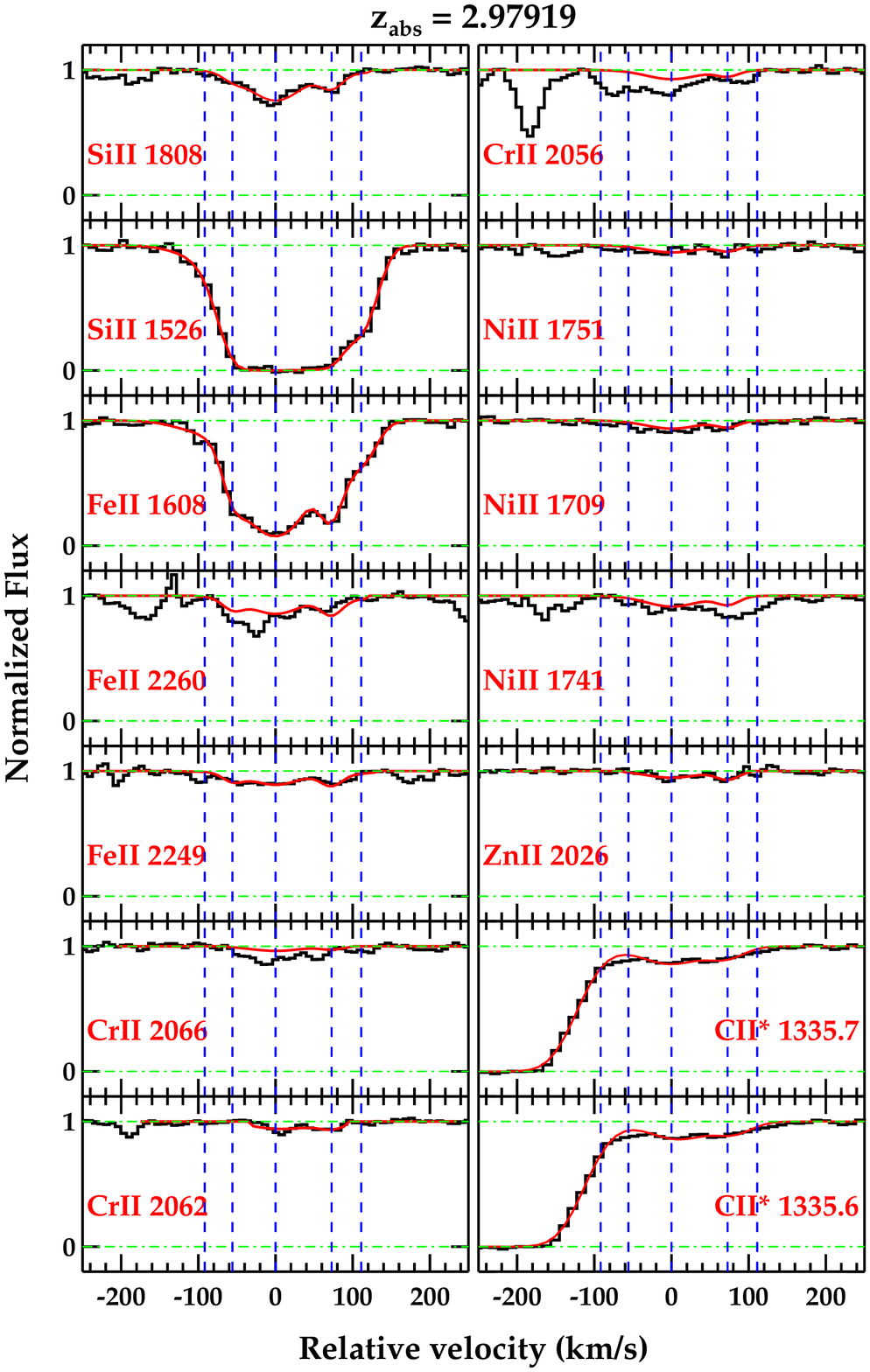} 
 }
\caption{Velocity plot of low ion absorption lines detected in \zabs=2.9791 DLA together with the best fitted Voigt profiles.}
 \label{vpfit2}
 \end{figure}

\begin{figure}
  \centering
 \vbox{
 \includegraphics[bb=30 17 600 790,width=1.0\linewidth,clip=true]{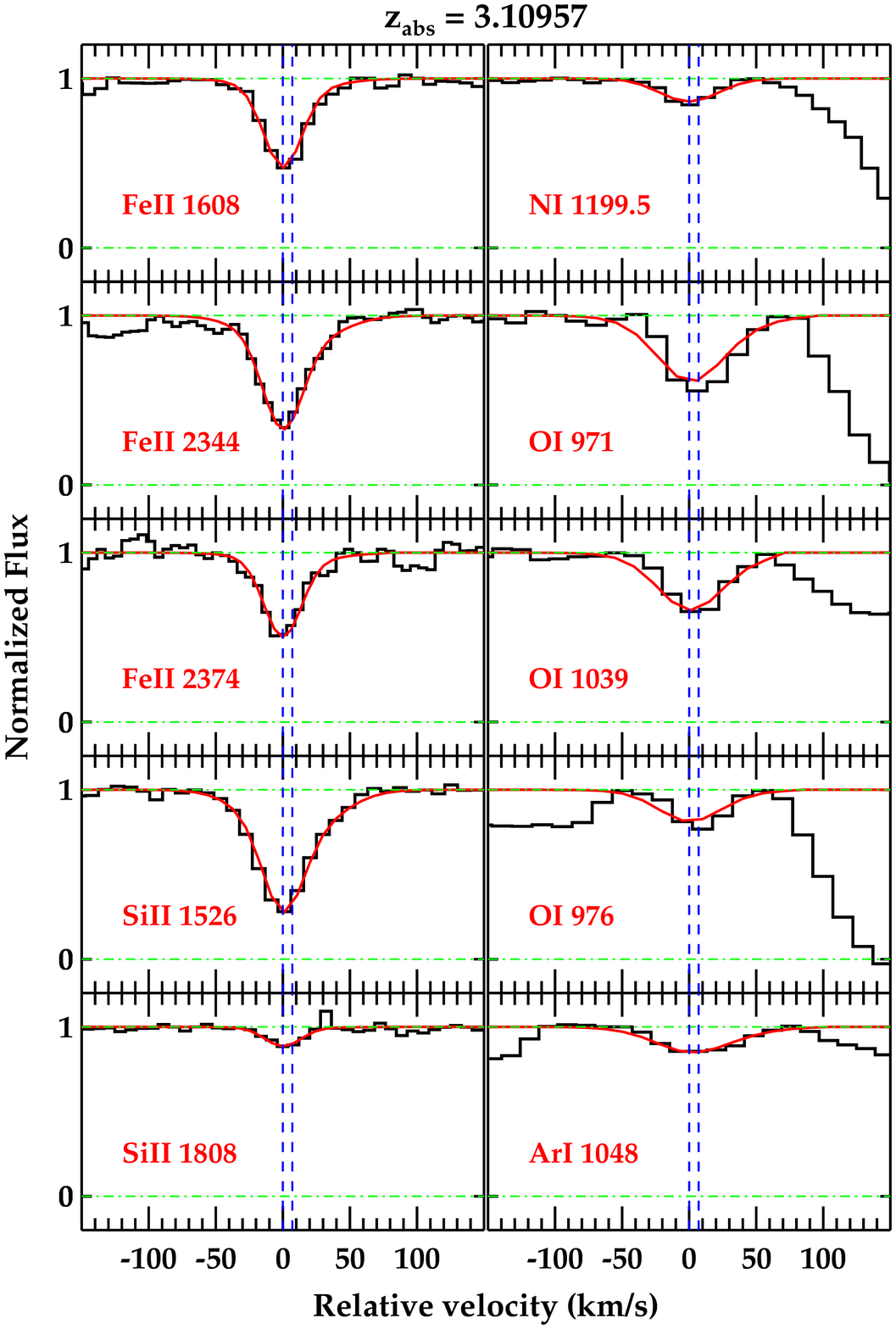} 
 }
\caption{Velocity plot of low ion absorption lines detected in \zabs=3.1095 DLA together with the best fitted Voigt profiles.}
 \label{vpfit3}
 \end{figure}

 \begin{figure}
  \centering
 \vbox{
 \includegraphics[bb=30 15 600 790,width=1.0\linewidth,clip=true]{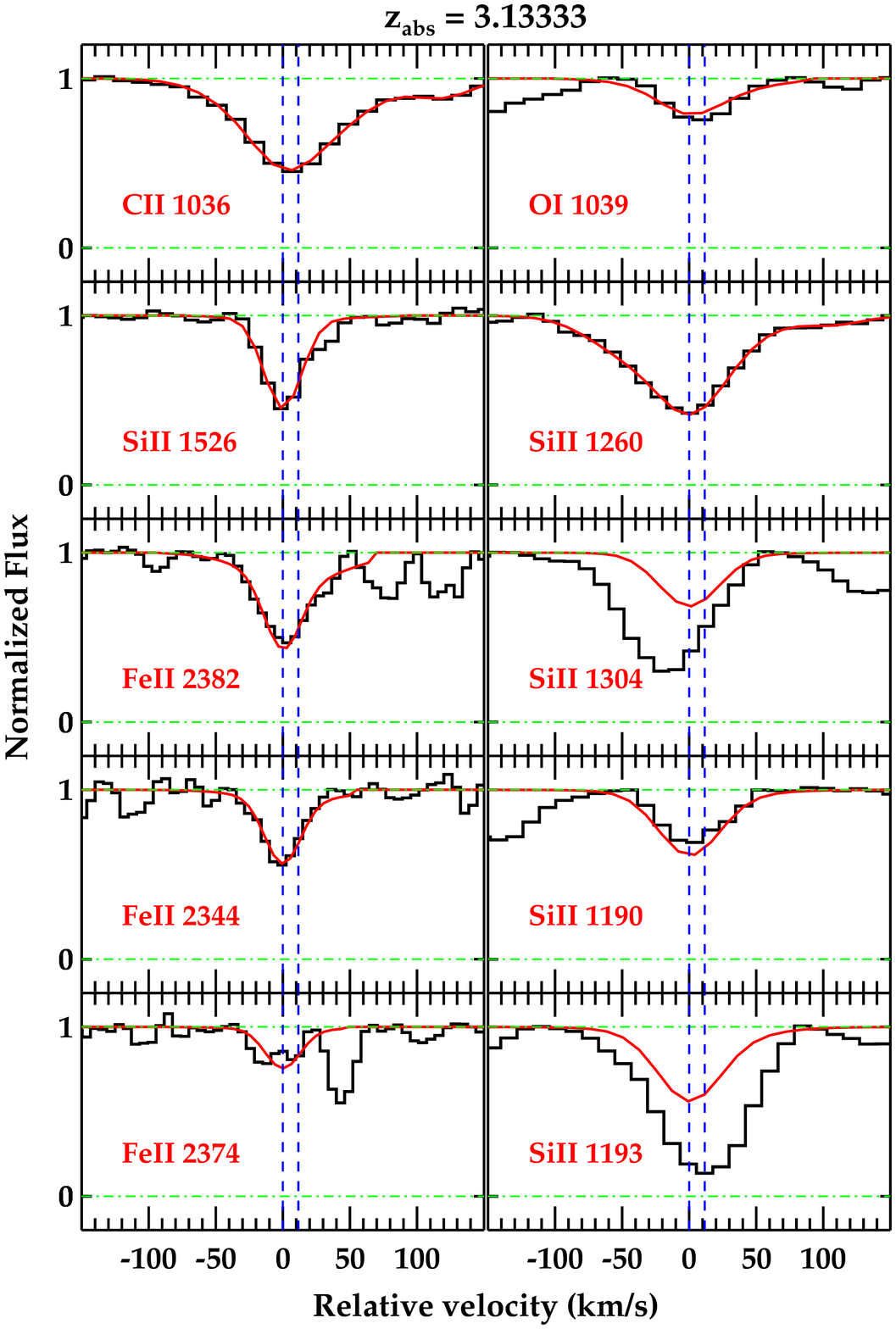}     
   }
 \caption{Velocity plot of low ion absorption lines detected in \zabs=3.1333 DLA together with the best fitted Voigt profiles.}
 \label{vpfit4}
 \end{figure} 
 \begin{figure}
  \centering
 { 
 {\includegraphics[bb=30 15 650 790,width=1.0\linewidth,clip=true]{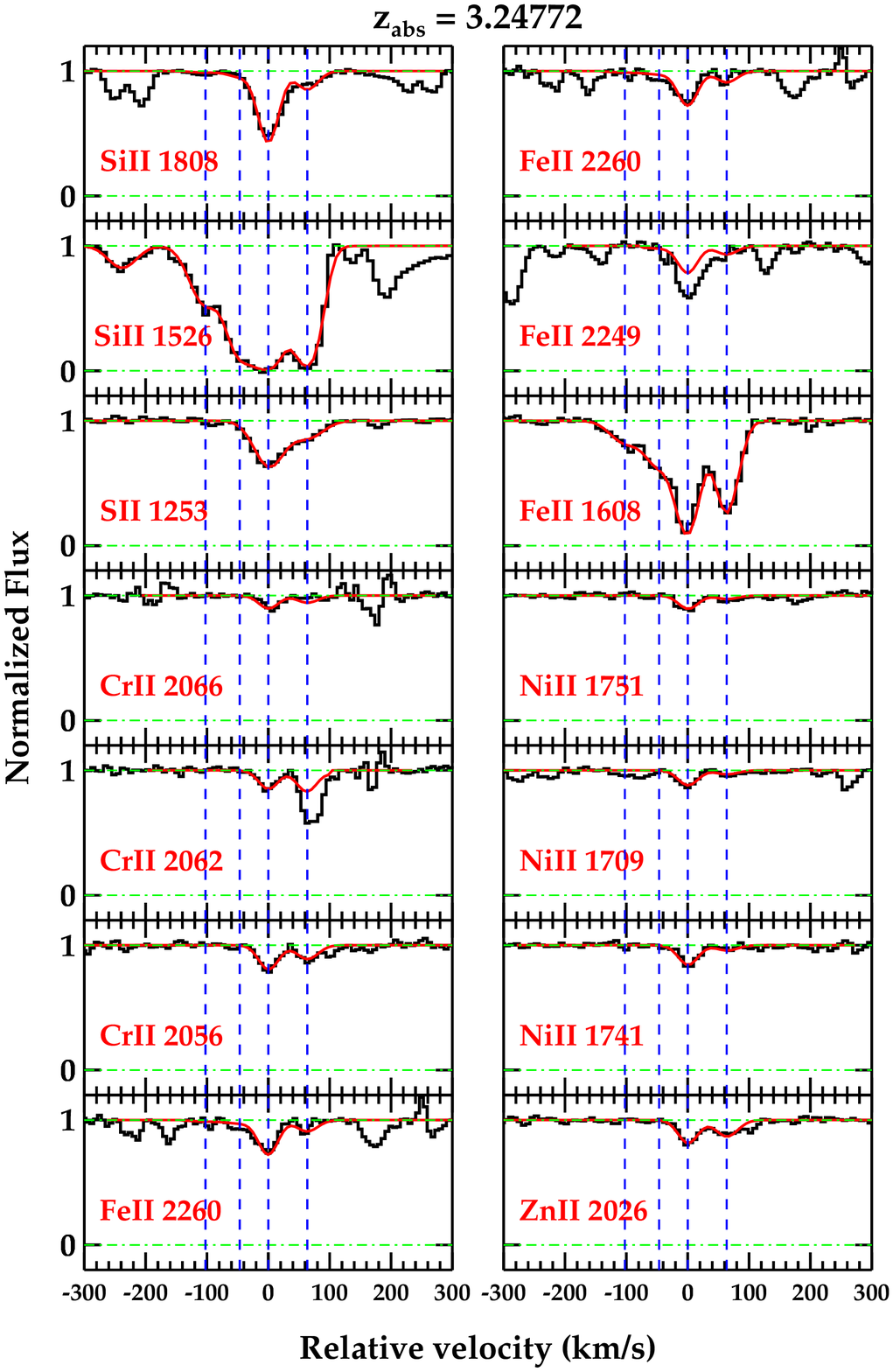}}
 }
 
 \caption{Velocity plot of low ion absorption lines detected in \zabs=3.2477 DLA together with the best fitted Voigt profiles.}
  \label{vpfit5}
 \end{figure} 


\begin{table}
 \centering
\caption{Voigt profile fit results for \zabs = 2.7377 system}
{%
\newcommand{\mc}[3]{\multicolumn{#1}{#2}{#3}}
\begin{tabular}{clllll}\hline \hline 
\mc{1}{l}{\zabs} & \mc{1}{c}{Species} & \mc{2}{c}{$\log N$ (\cmsq)} & \mc{2}{c}{$b$ (\kms)}\\ \hline 
\mc{1}{l}{2.737478} & \mc{1}{c}{\FeII} & \mc{2}{c}{ $12.37\pm0.36$} & \mc{2}{c}{ $3.54\pm1.26$}\\ 
\mc{1}{l}{2.737737} & \mc{1}{c}{\FeII} & \mc{2}{c}{$13.49\pm0.04$} & \mc{2}{c}{$6.65\pm1.29$}\\
\mc{1}{l}{} & \mc{1}{c}{\SiII } & \mc{2}{c}{$13.90\pm0.14$ } & \mc{2}{c}{  }\\
\mc{1}{l}{ } & \mc{1}{c}{ \SII} & \mc{2}{c}{$13.73\pm0.35$ } & \mc{2}{c}{  }\\  \hline
\end{tabular}
}%
\label{at1}
\end{table}

\begin{table}
 \centering
\caption{Voigt profile fit results for \zabs = 2.9791 system}
{%
\newcommand{\mc}[3]{\multicolumn{#1}{#2}{#3}}
\begin{tabular}{clllll}\hline \hline 
\mc{1}{l}{\zabs} & \mc{1}{c}{Species} & \mc{2}{c}{$\log N$ (\cmsq)} & \mc{2}{c}{$b$ (\kms)}\\ \hline 
\mc{1}{l}{ 2.978455} & \mc{1}{c}{\SiII } & \mc{2}{c}{$14.22\pm0.21$ } & \mc{2}{c}{$4.87\pm1.04$ }\\
\mc{1}{l}{ } & \mc{1}{c}{\FeII } & \mc{2}{c}{$14.34\pm0.11$ } & \mc{2}{c}{}\\

\mc{1}{l}{ 2.977978} & \mc{1}{c}{\SiII } & \mc{2}{c}{$13.21\pm0.15$ } & \mc{2}{c}{$46.58\pm2.83$ }\\
\mc{1}{l}{ } & \mc{1}{c}{\FeII } & \mc{2}{c}{$13.53\pm0.09$ } & \mc{2}{c}{ }\\

\mc{1}{l}{2.979195 } & \mc{1}{c}{\SiII} & \mc{2}{c}{$15.36\pm0.02$ } & \mc{2}{c}{$42.85\pm1.25$ }\\
\mc{1}{l}{  } & \mc{1}{c}{\FeII } & \mc{2}{c}{$14.94\pm0.02$ } & \mc{2}{c}{  }\\
\mc{1}{l}{  } & \mc{1}{c}{\ZnII } & \mc{2}{c}{$12.23\pm0.08$ } & \mc{2}{c}{  }\\
\mc{1}{l}{  } & \mc{1}{c}{\CrII } & \mc{2}{c}{$13.05\pm0.08$ } & \mc{2}{c}{  }\\
\mc{1}{l}{  } & \mc{1}{c}{\NiII } & \mc{2}{c}{$13.57\pm0.13$ } & \mc{2}{c}{  }\\
\mc{1}{l}{  } & \mc{1}{c}{\cstar } & \mc{2}{c}{$13.50\pm0.03$ } & \mc{2}{c}{  }\\

\mc{1}{l}{2.980159 } & \mc{1}{c}{\SiII } & \mc{2}{c}{ $14.74\pm0.07$} & \mc{2}{c}{$10.05\pm0.93$ }\\
\mc{1}{l}{   } & \mc{1}{c}{\FeII } & \mc{2}{c}{$14.61\pm0.06$ } & \mc{2}{c}{  }\\
\mc{1}{l}{   } & \mc{1}{c}{ \ZnII} & \mc{2}{c}{$11.98\pm0.13$ } & \mc{2}{c}{  }\\
\mc{1}{l}{   } & \mc{1}{c}{\CrII } & \mc{2}{c}{$12.51\pm0.20$ } & \mc{2}{c}{  }\\
\mc{1}{l}{   } & \mc{1}{c}{\NiII } & \mc{2}{c}{$13.13\pm0.29$ } & \mc{2}{c}{  }\\
\mc{1}{l}{   } & \mc{1}{c}{\cstar } & \mc{2}{c}{$13.10\pm0.07$} & \mc{2}{c}{  }\\ 

\mc{1}{l}{2.980669 } & \mc{1}{c}{\SiII } & \mc{2}{c}{$14.02\pm0.03$ } & \mc{2}{c}{$18.87\pm2.20$ }\\
\mc{1}{l}{ } & \mc{1}{c}{\FeII } & \mc{2}{c}{$13.91\pm0.04$ } & \mc{2}{c}{ }\\ \hline

\end{tabular}
}%
\label{at2}
\end{table}

\begin{table}
 \centering
\caption{Voigt profile fit results for \zabs = 3.1095 system}
{%
\newcommand{\mc}[3]{\multicolumn{#1}{#2}{#3}}
\begin{tabular}{clllll}\hline \hline 
\mc{1}{l}{\zabs} & \mc{1}{c}{Species} & \mc{2}{c}{$\log N$ (\cmsq)} & \mc{2}{c}{$b$ (\kms)}\\ \hline 
\mc{1}{l}{3.109576 } & \mc{1}{c}{\OI } & \mc{2}{c}{ $15.18\pm0.12$} & \mc{2}{c}{$6.13\pm0.51$ }\\
\mc{1}{l}{  } & \mc{1}{c}{\SiII } & \mc{2}{c}{ $14.62\pm0.07$} & \mc{2}{c}{  }\\
\mc{1}{l}{  } & \mc{1}{c}{\FeII } & \mc{2}{c}{$14.34\pm0.09$ } & \mc{2}{c}{  }\\
\mc{1}{l}{  } & \mc{1}{c}{\NI } & \mc{2}{c}{$13.39\pm0.10$} & \mc{2}{c}{  }\\
\mc{1}{l}{  } & \mc{1}{c}{\ArI } & \mc{2}{c}{ $12.76\pm0.43$} & \mc{2}{c}{  }\\
\mc{1}{l}{  } & \mc{1}{c}{\CII } & \mc{2}{c}{ $14.20\pm0.86$} & \mc{2}{c}{  }\\

\mc{1}{l}{3.109672 } & \mc{1}{c}{\OI } & \mc{2}{c}{$14.55\pm0.13$ } & \mc{2}{c}{$37.17\pm2.74$ }\\
\mc{1}{l}{  } & \mc{1}{c}{\SiII } & \mc{2}{c}{$13.65\pm0.06$ } & \mc{2}{c}{ }\\
\mc{1}{l}{  } & \mc{1}{c}{\FeII } & \mc{2}{c}{$13.33\pm0.07$ } & \mc{2}{c}{ }\\
\mc{1}{l}{  } & \mc{1}{c}{\ArI } & \mc{2}{c}{$13.05\pm0.22$ } & \mc{2}{c}{ }\\
\mc{1}{l}{  } & \mc{1}{c}{\CII } & \mc{2}{c}{$14.43\pm0.06$ } & \mc{2}{c}{ }\\ \hline
\end{tabular}
}%
\label{at3}
\end{table}

\begin{table}
 \centering
\caption{Voigt profile fit results for \zabs = 3.1333}
{%
\newcommand{\mc}[3]{\multicolumn{#1}{#2}{#3}}
\begin{tabular}{clllll}\hline \hline 
\mc{1}{l}{\zabs} & \mc{1}{c}{Species} & \mc{2}{c}{$\log N$ (\cmsq)} & \mc{2}{c}{$b$ (\kms)}\\ \hline 
\mc{1}{l}{3.133331 } & \mc{1}{c}{\FeII } & \mc{2}{c}{$13.77\pm0.14$ } & \mc{2}{c}{$4.71\pm0.72$ }\\
\mc{1}{l}{  } & \mc{1}{c}{\SiII } & \mc{2}{c}{$14.66\pm0.44$ } & \mc{2}{c}{  }\\
\mc{1}{l}{  } & \mc{1}{c}{\CII } & \mc{2}{c}{$14.39\pm0.69$ } & \mc{2}{c}{  }\\
\mc{1}{l}{  } & \mc{1}{c}{\OI} & \mc{2}{c}{ $14.70\pm0.21$} & \mc{2}{c}{  }\\ 
\mc{1}{l}{3.133491 } & \mc{1}{c}{\FeII } & \mc{2}{c}{ $12.89\pm0.09$} & \mc{2}{c}{ $47.59\pm6.46$}\\
\mc{1}{l}{  } & \mc{1}{c}{\SiII } & \mc{2}{c}{ $12.61\pm0.14$} & \mc{2}{c}{  }\\
\mc{1}{l}{  } & \mc{1}{c}{ \CII} & \mc{2}{c}{$14.19\pm0.05$ } & \mc{2}{c}{  }\\
\mc{1}{l}{  } & \mc{1}{c}{\OI } & \mc{2}{c}{$14.55\pm0.16$ } & \mc{2}{c}{  }\\ \hline

\end{tabular}
}%
\label{at4}
\end{table}

\begin{table}
 \centering
\caption{Voigt profile fit results for \zabs = 3.2477 system}
{%
\newcommand{\mc}[3]{\multicolumn{#1}{#2}{#3}}
\begin{tabular}{clllll}\hline \hline 
\mc{1}{l}{\zabs} & \mc{1}{c}{Species} & \mc{2}{c}{$\log N$ (\cmsq)} & \mc{2}{c}{$b$ (\kms)}\\ \hline 
\mc{1}{l}{ 3.246272} & \mc{1}{c}{ \FeII} & \mc{2}{c}{$13.65\pm0.11$ } & \mc{2}{c}{$27.79\pm6.35$ }\\
\mc{1}{l}{ } & \mc{1}{c}{\SiII } & \mc{2}{c}{$13.85\pm0.08$ } & \mc{2}{c}{ }\\
\mc{1}{l}{ 3.247061} & \mc{1}{c}{\FeII } & \mc{2}{c}{$14.01\pm0.06$ } & \mc{2}{c}{$18.88\pm5.44$ }\\
\mc{1}{l}{ } & \mc{1}{c}{\SiII } & \mc{2}{c}{$14.33\pm0.07$ } & \mc{2}{c}{ } \\ 
\mc{1}{l}{ 3.247724} & \mc{1}{c}{\FeII } & \mc{2}{c}{$14.95\pm0.05$ } & \mc{2}{c}{$13.63\pm0.86$ }\\
\mc{1}{l}{  } & \mc{1}{c}{\SiII } & \mc{2}{c}{ $15.59\pm0.02$} & \mc{2}{c}{  }\\
\mc{1}{l}{  } & \mc{1}{c}{\SII } & \mc{2}{c}{$14.98\pm0.04$ } & \mc{2}{c}{  }\\
\mc{1}{l}{  } & \mc{1}{c}{\CrII } & \mc{2}{c}{$13.19\pm0.05$ } & \mc{2}{c}{  }\\
\mc{1}{l}{  } & \mc{1}{c}{\NiII } & \mc{2}{c}{$13.54\pm0.05$ } & \mc{2}{c}{  }\\
\mc{1}{l}{  } & \mc{1}{c}{\ZnII} & \mc{2}{c}{$12.47\pm0.06$ } & \mc{2}{c}{  }\\
\mc{1}{l}{3.248624} & \mc{1}{c}{\FeII } & \mc{2}{c}{$14.44\pm0.03$ } & \mc{2}{c}{ $16.34\pm0.89$}\\
\mc{1}{l}{  } & \mc{1}{c}{ \SiII} & \mc{2}{c}{$14.84\pm0.06$ } & \mc{2}{c}{  }\\
\mc{1}{l}{  } & \mc{1}{c}{\SII } & \mc{2}{c}{$14.41\pm0.07$ } & \mc{2}{c}{  }\\
\mc{1}{l}{  } & \mc{1}{c}{\CrII} & \mc{2}{c}{$12.95\pm0.08$ } & \mc{2}{c}{  }\\
\mc{1}{l}{  } & \mc{1}{c}{ \NiII} & \mc{2}{c}{ $12.97\pm0.17$} & \mc{2}{c}{  }\\
\mc{1}{l}{  } & \mc{1}{c}{\ZnII } & \mc{2}{c}{$12.34\pm0.07$ } & \mc{2}{c}{  }\\ \hline  

\end{tabular}
}%
\label{at5}
\end{table}

\end{document}